\documentclass[aps,prd,twoside,twocolumn,superscriptaddress,floatfix,nofootinbib,showpacs]{revtex4-1}
\usepackage{amsmath}
\usepackage{bm}
\usepackage{xcolor}
\usepackage{hyperref}
\usepackage{url}
\usepackage{graphicx}
\usepackage[large]{subfigure}
\usepackage{amssymb}
\usepackage[amssymb]{SIunits}
\usepackage{aas_macros}
\usepackage{natbib}

\usepackage{multirow}

\newcommand\beqn{\begin{eqnarray}}
\newcommand\eeqn{\end{eqnarray}}

\usepackage{environ}
\NewEnviron{myequation}{%
\begin{equation}
\scalebox{0.9}{$\BODY$}
\end{equation}
}

\newcommand*{\field}[1]{\mathbb{#1}}

\newcommand\lsim{\mathrel{\rlap{\lower4pt\hbox{\hskip1pt$\sim$}}
        \raise1pt\hbox{$<$}}}
\newcommand\gsim{\mathrel{\rlap{\lower4pt\hbox{\hskip1pt$\sim$}}
        \raise1pt\hbox{$>$}}}

\usepackage[mathscr]{eucal}	
\DeclareTextSymbol{\degre}{T1}{23}

\newcommand{\vl} { {\mbf{\ell}} }
\newcommand{\vL} { {\mbf{L}} }

\newcommand{\vd}{{\mbf{d}}}

\newcommand{\n} {\hat{\mbf{n}}}

\newcommand{\beq} {\begin{equation}}
\newcommand{\eeq} {\end{equation}}
\newcommand{\bal} {\begin{aligned}}
\newcommand{\eal} {\end{aligned}}

\newcommand{\mbf}[1]{\mbox{\boldmath$#1$}}

\begin{document}

\title{Foreground-immune CMB lensing with shear-only reconstruction}

\author{Emmanuel Schaan}
\email{eschaan@lbl.gov}
\affiliation{Lawrence Berkeley National Laboratory, One Cyclotron Road, Berkeley, CA 94720, USA}
\affiliation{Berkeley Center for Cosmological Physics, University of California, Berkeley, CA 94720, USA}
\author{Simone Ferraro}
\email{sferraro@lbl.gov}
\affiliation{Lawrence Berkeley National Laboratory, One Cyclotron Road, Berkeley, CA 94720, USA}
\affiliation{Berkeley Center for Cosmological Physics, University of California, Berkeley, CA 94720, USA}

\begin{abstract} 

CMB lensing from current and upcoming wide-field CMB experiments such as AdvACT, SPT-3G 
and Simons Observatory
relies heavily on temperature
(vs. polarization).
In this regime, foreground contamination to the temperature map produces significant lensing biases,
which cannot be fully controlled by multi-frequency component separation, masking or bias hardening.

In this letter, we split the standard CMB lensing quadratic estimator into a new set of optimal `multipole' estimators.
On large scales, these multipole estimators reduce to the known magnification and shear estimators, and a new shear B-mode estimator.
We leverage the different symmetries of the lensed CMB and extragalactic foregrounds to argue that the shear-only estimator should be approximately immune to extragalactic foregrounds.
We build a new method to compute
separately 
and without noise
the primary, secondary and trispectrum biases to CMB lensing from foreground simulations. 
Using this method,
we demonstrate that the shear estimator is
indeed
 insensitive to extragalactic foregrounds,
even when applied to a single-frequency temperature map contaminated with CIB, tSZ, kSZ and radio point sources. 
This 
dramatic reduction in foreground biases
allows us to include higher temperature multipoles than with the standard quadratic estimator, 
thus increasing the total lensing signal-to-noise beyond the quadratic estimator.
In addition, magnification-only and shear B-mode estimators provide useful diagnostics for potential residuals.

Our python code \texttt{LensQuEst} to forecast the signal-to-noise of the various estimators, generate mock maps, lense them, and apply the various lensing estimators to them is publicly available at
\url{https://github.com/EmmanuelSchaan/LensQuEst}.

\end{abstract}

\maketitle

\section{Introduction}

Weak lensing of the CMB measures the projected matter distribution throughout the observable Universe, and is one of the most promising probes of dark energy, modified gravity and neutrino masses \cite{2006PhR...429....1L, 2010GReGr..42.2197H}.
As the measurement precision increases, systematic biases become more important.  
While CMB-S4 \cite{2016arXiv161002743A} lensing data should be polarization-dominated in the future,
in the coming decade, CMB lensing measurements from AdvACT \cite{2016JLTP..184..772H}, SPT-3G \cite{2014SPIE.9153E..1PB} 
and Simons Observatory \cite{2018arXiv180807445T}
will rely heavily on temperature. 
In this regime, extragalactic foregrounds such as 
the cosmic infrared background (CIB),
the thermal Sunyaev-Zel'dovich effect (tSZ),
the kinematic Sunyaev-Zel'dovich effect (kSZ)
and radio point sources (PS)
 can produce biases much larger than the statistical errors, if unaccounted for
 \cite{2014ApJ...786...13V, 2014JCAP...03..024O, 2018arXiv180208230M, 2018PhRvD..97b3512F}.
Mitigation methods have been proposed.
For example, masking individually detected or know sources can decrease the bias, and techniques such as bias hardening \cite{2014JCAP...03..024O, 2013MNRAS.431..609N} are effective when the foreground trispectrum is known.
Multi-frequency component separation \cite{2018arXiv180208230M} can reduce or null specific foregrounds components. 
However, a minimum-variance multifrequency analysis only leads to a modest reduction in foregrounds, and simultaneously nulling tSZ and CIB comes at a large cost, increasing the noise power spectrum by a factor as large as $50$  \cite{2018arXiv180807445T}. 
Furthermore, multi-frequency component separation has no effect on the kSZ, which alone causes a significant lensing bias \cite{2018PhRvD..97b3512F}.
New methods are therefore needed in order to produce unbiased lensing measurements from CMB temperature maps.

In this letter, we explore a new approach, leveraging the differing symmetries of the lensing deflections and extragalactic foregrounds in order to separate them. 
Indeed, as we argue below, extragalactic foregrounds are degenerate with lensing magnification (local monopole distortion of the power spectrum), but not with lensing shear (local quadrupolar distortion) or higher order multipoles.
Throughout this letter, we consider lensing measurements from CMB temperature only, rather than polarization, although we expect a similar approach to work in polarization too.

\section{Lensing multipole estimators}

\subsection*{Estimators}

Weak lensing modulates the 2d CMB power spectrum, creating local distortions. These distortions to the power spectrum can be decomposed into a monopole ($m=0$) corresponding to an isotropic magnification or demagnification, a quadrupole ($m=2$) corresponding to shearing, as well as higher order even multipoles. Mathematically, the presence of a fixed lensing convergence $\kappa_{\vL}$, creates off-diagonal correlations in the observed CMB temperature $T$:
\begin{equation}
\langle T_{\vl + \frac{\vL}{2}} T_{\frac{\vL}{2} - \vl} \rangle
=
f_{\vl + \frac{\vL}{2} ,  \frac{\vL}{2} - \vl}^\kappa \;
\kappa_{\vL} 
+\mathcal{O}(\kappa^2).\\
\label{eq:kapparesponse} 
\end{equation}
The angular dependence of the response function $f^{\kappa}$ can be expanded in multipoles of the angle $\theta_{\vL, \vl}$ between $\vl$ and $\vL$:
\beq
f_{\vl + \frac{\vL}{2} ,  \frac{\vL}{2} - \vl}^\kappa
=
\sum_{m \text{ even}}
f_{L, \ell}^m
\cos \left( m \theta_{\vL, \vl} \right),
\eeq
which defines the $m-$th multipole response function $f_{L, \ell}^m$. These can be used in Eq. \ref{eq:kapparesponse} to obtain an estimator of $\kappa_{\vL}$, from multipole $m$ only.
Explicit minimum variance expressions are given in the Supplemental Material, and 
Fig. \ref{fig:multipole_lensing} shows that the monopole and quadrupole estimators contain most of the lensing signal-to-noise, allowing us to neglect estimators with $m>2$ in practice.
\begin{figure}[ht]
\centering
\includegraphics[width=1\columnwidth]{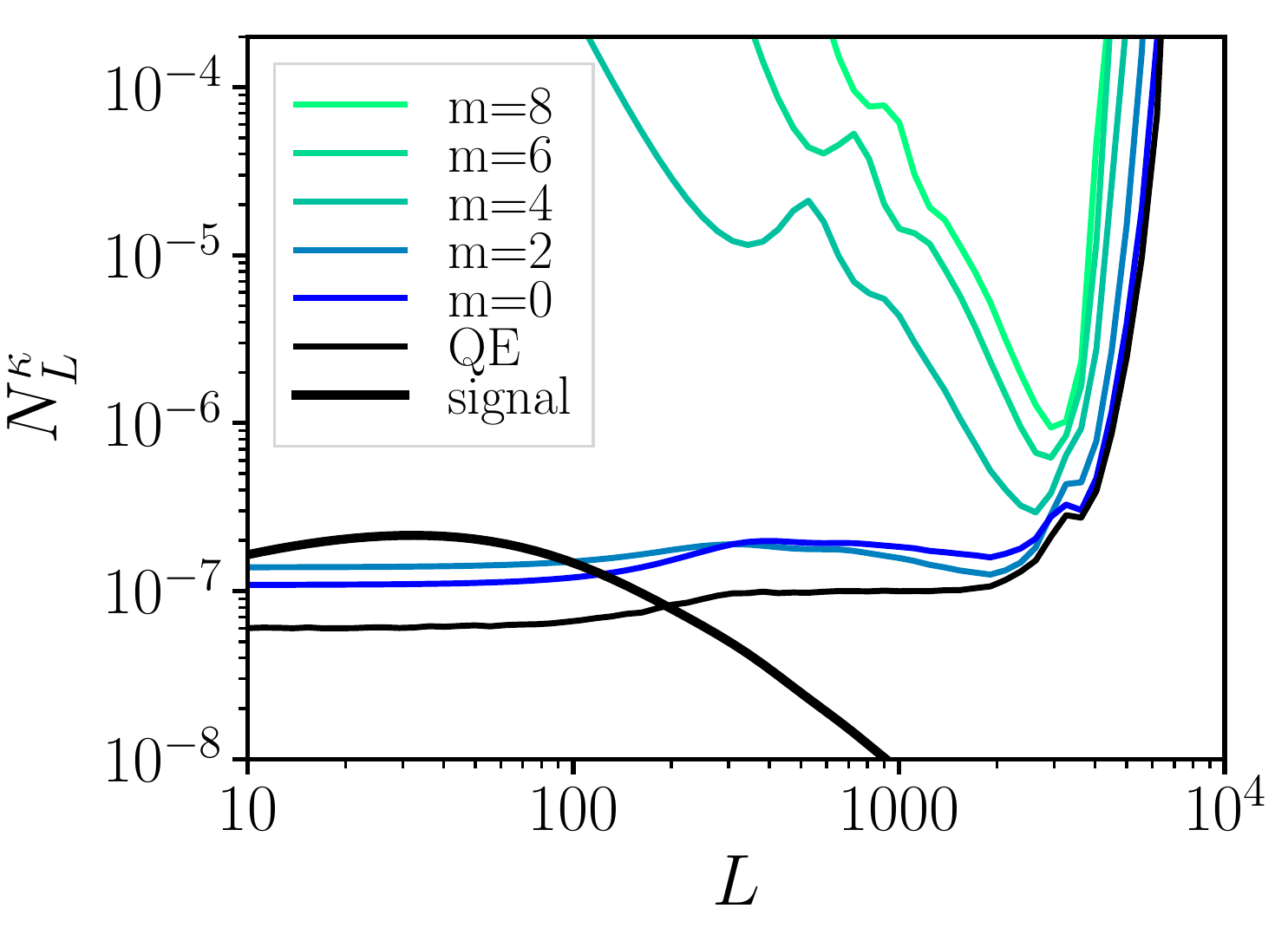}
\caption{Noise power spectrum of the lensing convergence $\kappa$, reconstructed with the optimal quadratic multipole estimators. 
Monopole ($m=0$) and quadrupole ($m=2$) estimators contain most of the lensing signal-to-noise.
The multipole estimators are uncorrelated for $L \lesssim 300$.
}
\label{fig:multipole_lensing}
\end{figure}
To allow a fast evaluation with FFT, we can replace these non-separable optimal multipole estimators by their limits in the `large-scale lens regime', where large-scale ($L \lesssim 300$) lensing modes are reconstructed from small-scale ($\ell \gtrsim 300$) temperature modes.
In this regime, our optimal monopole and quadrupole estimators reduce to the magnification\footnote{To be consistent with the optical lensing literature, this estimator should be called `convergence' instead of `magnification'. Since we already use the name `convergence' to designate the lensing field $\kappa$ that is being reconstructed, we decided to call shear and magnification the two distinct effects, to avoid confusion.} 
and shear E-mode estimators of  \cite{2008MNRAS.388.1819L, 2012PhRvD..85d3016B, 2017arXiv170902227P} (see also \cite{1999PhRvD..59l3507Z, 2004NewA....9..417P, 2008MNRAS.388.1819L, 2010PhRvD..81l3015L}),
as well as a new shear B-mode estimator:
\beq
\kappa_\vL
=
\frac{
\int \frac{d ^2 \vl}{(2\pi)^2}\;
T_{\vl } T_{\vL-\vl}\;
g_{\vL, \vl}
}
{
\frac{2\vL}{L^2} 
\cdot
\int \frac{d ^2 \vl}{(2\pi)^2}\;
g_{\vL, \vl}
\left[
\vl C^0_{\ell}
+
\left( {\vL-\vl} \right)C^0_{\vL-\vl}
\right]
}
,
\label{eq:estimators}
\eeq
where
\beq
\left\{
\bal
&g_{\vL, \vl}^\text{magnification}
=
\frac{C^0_\ell}{2(C^\text{total}_\ell)^2}
\frac{d \ln \ell^2 C^0_\ell}{d \ln \ell},\\
&g_{\vL, \vl}^\text{shear E}
=
\text{cos}(2\theta_{\vL, \vl})
\frac{C^0_\ell}{2(C^\text{total}_\ell)^2}
\frac{d \ln C^0_\ell}{d \ln \ell},\\
&g_{\vL, \vl}^\text{shear B}
=
\text{sin}(2\theta_{\vL, \vl})
\frac{C^0_\ell}{2(C^\text{total}_\ell)^2}
\frac{d \ln C^0_\ell}{d \ln \ell}.
\eal
\right.
\label{eq:weights}
\eeq
These estimators should only be interpreted as measuring magnification and shear in the large-scale lens regime ($L \ll \ell$).
However, they remain unbiased lensing estimators on all scales.
They match the harmonic-space version of \cite{2012PhRvD..85d3016B, 2017arXiv170902227P}, 
after normalizing them to be unbiased and with the substitution $T_{\vl + \vL/2 } T_{\vL/2-\vl} \rightarrow T_{\vl} T_{\vL-\vl}$ to allow fast evaluation with FFT.
We further substitute the lensed CMB power spectrum to $C^0$, as is customary for the QE \cite{2011PhRvD..83d3005H, 2011JCAP...03..018L}.
As shown in the Suppl. Mat. Fig.~1, the magnification and shear estimators are optimal on large scales ($L \lesssim 300$), where they have the same noise as the optimal $m=0$ and $m=2$ estimators, are roughly uncorrelated, and recover the signal-to-noise of the standard quadratic estimator (QE).
In the Born approximation, the shear $B$-mode estimator has zero response to lensing and provides a useful null test.
As we show below, it also allows us to detect and subtract any potential `secondary foreground bias' (defined below).

\subsection*{Statistical signal-to-noise}

Throughout this letter, we consider an upcoming stage 3 (`CMB S3') experiment, with $1.4'$ beam FWHM and $7 \mu K'$ sensitivity at 148GHz.
We apply the lensing estimators to the single-frequency map at 148GHz, without any multi-frequency component separation.
For the lensing weights, we include the lensed CMB, all the foregrounds of Sec.~\ref{sec:simluations} and the detector white noise in the total power spectrum.

\begin{figure}[h]
\centering
\includegraphics[width=0.9\columnwidth]{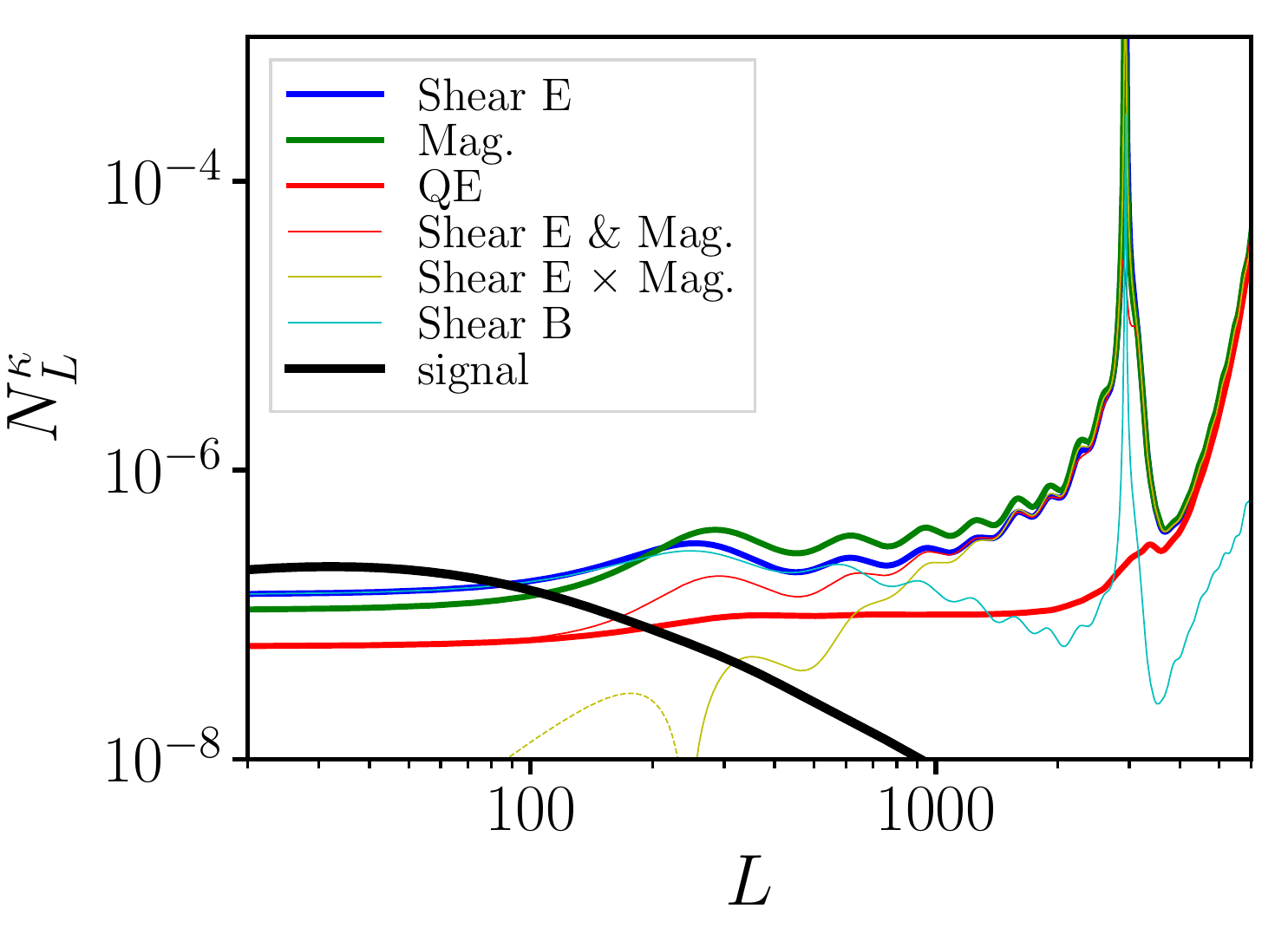}
\caption{
Lensing reconstruction noise per lensing multipole for the standard quadratic estimator (QE, red), the magnification (green), shear E-mode (blue) and B-mode (cyan) estimators, when using temperature modes $\ell=30-3500$. 
Below multipoles of a few hundred, the shear E and magnification estimators are roughly uncorrelated, and recover the QE when combined, taking into account their noise covariance. 
Shear E and shear B have similar noise for low multipoles, which makes the shear B a useful null test to compare to shear E.
}
\label{fig:summary_snr_qsd1}
\end{figure}

\begin{figure}[h]
\centering
\includegraphics[width=0.9\columnwidth]{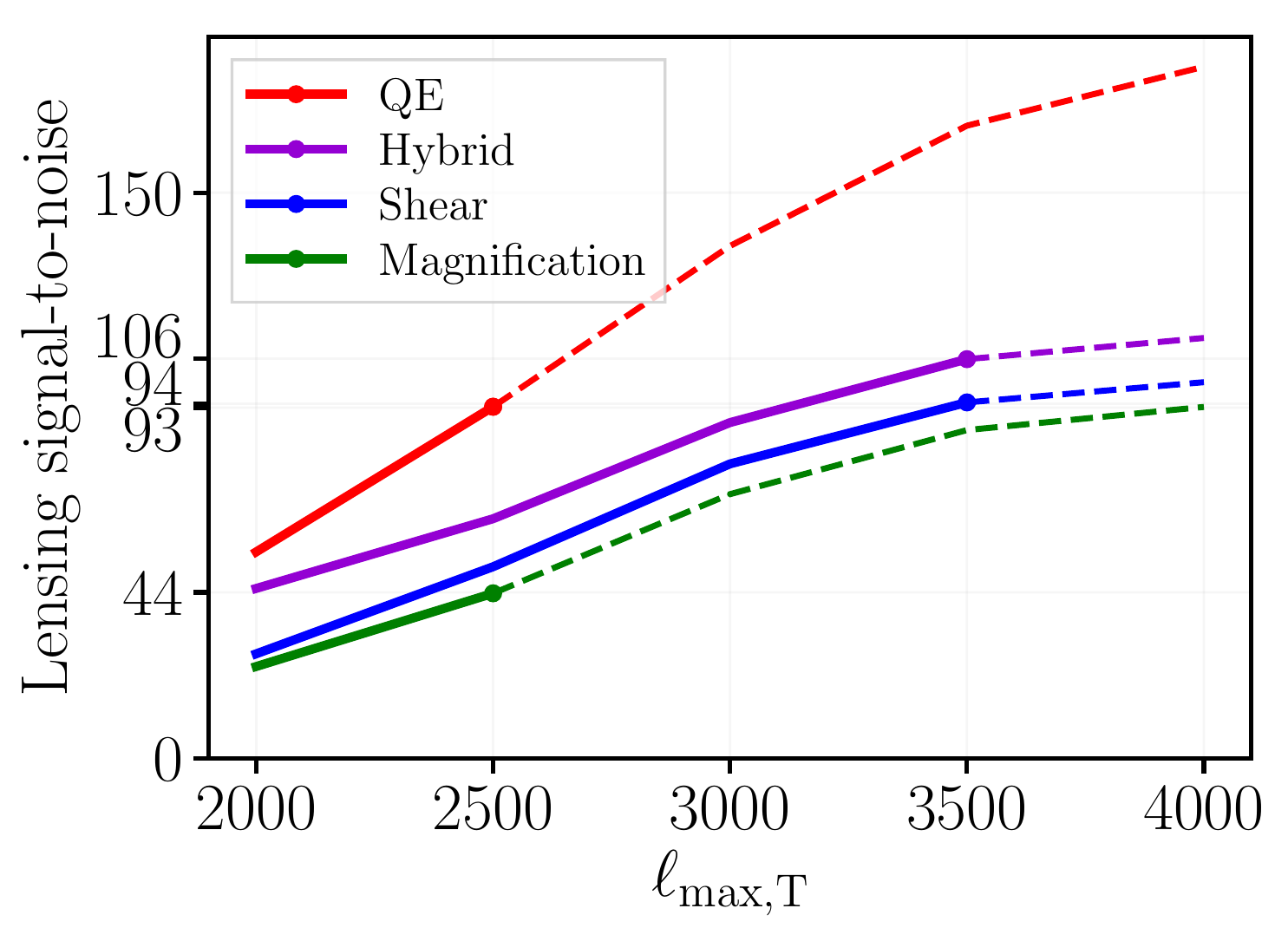}
\caption{
Total signal-to-noise on the amplitude of the lensing power spectrum, including cosmic variance, as a function of the maximum temperature multipole $\ell_{\text{max}, T}$, for $f_\text{sky}=1$. Different colors correspond to the different estimators. 
Dashed lines indicate when foreground biases are larger than the statistical uncertainty, even after masking point sources detected at 5$\sigma$.
At fixed $\ell_{\text{max}, T}$, the signal-to-noise in either shear or magnification is about $60\%$ of the signal-to-noise of the QE.
However, as we show below, keeping the foreground bias below the statistical error requires $\ell_{\text{max}, T}=2500$ for the QE (red dot, $S/N=70$), compared to $\ell_{\text{max}, T}=3500$ for the shear estimator (blue dot, $S/N=77$): hence the final shear signal-to-noise exceeds that of the QE by $10\%$.
A hybrid estimator QE($\ell\leq2000$) \& shear($\ell=2000-3500$) is shown in purple, and increases the signal-to-noise by $37\%$ compared to the standard QE($\ell\leq 2500$).}
\label{fig:summary_snr_qsd2}
\end{figure}

Intuitively, Eq.~\eqref{eq:weights} means that magnification can only be measured from a non-scale-invariant power spectrum ($d \ln \ell^2C^0_\ell/ d \ln \ell \neq 0$), and shear only from a non-white power spectrum ($d \ln C^0_\ell/d \ln \ell \neq 0$).
The unlensed CMB power spectrum is neither scale-invariant nor white, so a similar signal-to-noise is expected for the shear and magnification estimators.
Indeed, as shown in Fig.~\ref{fig:summary_snr_qsd1}, the lensing noise in shear and magnification is comparable.
This is convenient: shear and magnification estimators can be compared as a consistency check for residual foregrounds.
At fixed $\ell_{\text{max},T}$, the total signal-to-noise in either shear or magnification is about $60\%$ of that in the QE, including the cosmic variance.
However, as we show below, the shear estimator is less affected by foregrounds, allowing to use $\ell_{\text{max},T}=3500$ instead of $\ell_{\text{max},T}=2500$ for the QE.
This allows to recover all of the signal-to-noise lost by discarding the magnification part.
To optimize further, we build a `hybrid estimator' by forming the minimum-variance linear combination of the magnification measured from $\ell_{\text{max},T}=2000$ (where foreground contamination is small) and the shear measured from $\ell_T=30-3500$. This minimum-variance linear combination takes into account the correlation between the estimators.
This `hybrid' estimator, shown in Fig.~\ref{fig:summary_snr_qsd2}, increases the SNR on the amplitude of lensing by $14\%$ compared to the QE with $\ell_{\text{max},T}=2500$, from 93 to 106.
A similar hybrid estimator, constructed from the multipole estimators rather than from the magnification and shear, will increase the SNR even further.

A spike in the noise power spectrum can be seen for the magnification and shear estimators in Fig.~\ref{fig:summary_snr_qsd1}, but not for the multipole estimators in 
Fig.~\ref{fig:multipole_lensing}.
This is a result of the approximate lensing weights in Eq.~\eqref{eq:weights}, only valid in the large-scale lens regime, which cause these estimators to have zero response to lensing (and thus infinite noise) at the location of the spike.

\subsection*{Expected sensitivity to foregrounds}
\label{sec:expected_sensitivity_to_foregrounds}

Extragalactic foregrounds dominate the lensed CMB on small scales ($\ell \gtrsim 3000$), where they are well described by a one-halo or shot noise term,
i.e. by a set of unclustered emission profiles (e.g., halos) or point sources (e.g., galaxies inside azimuthally-symmetric halos).
If the emission profiles are azimuthally-symmetric, the local foreground power spectrum on a small patch of the sky is isotropic, i.e. function of $ \ell  = |\vl|$ instead of $\vl$.
As a result, the corresponding foreground component modifies the observed power spectrum monopole ($m=0$), but not its higher multipoles.
This should bias the magnification estimator, and therefore the QE, but not the shear estimator.

If the foreground sources are halos with random independent ellipticities, or are point-like but clustered in elliptical filaments with random orientations, they produce extra noise in the shear estimator, analogous to the shape noise in galaxy lensing.
On the other hand, if the ellipticities of foreground halos or of their clustering (filaments) are aligned with the local tidal field, they will produce a bias to the shear estimator, analogously to intrinsic alignments in galaxy lensing (see App. D in \cite{2018arXiv180304975F}).

In summary, any extragalactic foreground biases the magnification estimator and the QE, whereas only foregrounds with specific anisotropies (intrinsic alignments) affect the shear estimators.
In the next section, we test this intuition with realistic foreground simulations.

\section{Sensitivity to foregrounds: simulations}
\label{sec:simluations}

\subsection*{Method}

We use simulated maps of lensing convergence, CIB, tSZ, kSZ and radio PS at 148GHz from \cite{2010ApJ...709..920S},
obtained by painting polytropic baryonic profiles on a large-box ($L = 1$ Gpc/$h$) N-body simulation.
Crucially, the gas density and temperature profiles given to a halo are not spherical, but instead follow the triaxiality of the local matter tidal tensor at the position of the halo.
As a result, these simulations include a reasonable level of `shape noise' and `intrinsic alignment'.
A halo catalog from this N-body simulation is also available. 
We re-weight these halos to match the redshift distribution of the LSST gold sample, with $i$-band magnitude $i < 25.3$ \cite{2009arXiv0912.0201L}
($dn/dz \propto (z/z_0)^2 e^{-z/z_0} / (2 z_0)$ with $z_0=0.24$),
and obtain a projected `galaxy' number density map $\delta_g$.
The `galaxy bias' measured from this map roughly matches the expected value $b(z) = 1 + 0.84 z$ \cite{2009arXiv0912.0201L}.
These maps have two crucial features: they are realistically correlated with each other, and have a reasonable level of non-Gaussianity.
The simulations also include the effect of anisotropic clustering of halos inside filaments,
of anisotropic halo profiles, including possible intrinsic alignments.
Our goal is to compute the foreground biases to the cross-correlation of CMB lensing with galaxies $C_L^{\kappa \delta_g}$ and to the CMB lensing auto-spectrum $C_L^{\kappa \kappa}$.

We subtract the mean emission in each foreground map, then rescale the maps by factors of order one to match the power spectrum model of \cite{2013JCAP...07..025D} (0.38 for CIB, 0.7 for tSZ, 0.82 for kSZ, 1.1 for radio PS).
Following \cite{2014ApJ...786...13V}, we then mask the point sources with flux $\gtrsim 5$mJy in each foreground map. \
To do so, we match-filtered the foreground maps with a profile corresponding to the beam and a noise determined by the total power spectrum (lensed CMB plus all foregrounds).
The resulting foreground power spectra are shown in the Suppl. Mat. Fig~3.

In principle, one should add all the foreground maps together to get the total bias, including their correct cross-correlations. 
However, component separation will reduce each foreground differently. For this reason, we analyze each foreground map separately.
This should allow the reader to quantify the foreground bias for any component separation method by rescaling our values appropriately.
In what follows, our lens reconstruction relies on temperature multipoles $\ell=30-3500$.
To measure the lensing bias due to the foregrounds,
we decompose the observed sky temperature $T_{\rm obs}$ into the lensed primary CMB $T_{\rm CMB}$, the foregrounds $T_f$ and the detector noise $T_\text{noise}$:
$T_{\rm obs} = T_{\rm CMB} + T_f + T_\text{noise} $.
We write $Q[T_A, T_B]$ for any quadratic estimator (QE, shear or magnification) applied to maps $T_A$ and $T_B$, symmetrized in $A\leftrightarrow B$.

As shown in \cite{2018PhRvD..97b3512F, 2014ApJ...786...13V, 2014JCAP...03..024O}, 
biases to the CMB lensing auto power spectrum $C_L^{\kappa \kappa}$ arise from the foreground \textit{bispectrum} (`primary' and `secondary' terms \cite{2014JCAP...03..024O}), and from the foreground \textit{trispectrum}.
We evaluate them as follows:

1) The \textit{primary bispectrum} term is computed as $2\langle Q[T_f, T_f] \; \kappa_\text{CMB} \rangle$, as in \cite{2014ApJ...786...13V, 2014JCAP...03..024O, 2018PhRvD..97b3512F}.

2) The \textit{secondary bispectrum} could in principle be computed as $4\langle Q[T_f, T_\text{CMB}] \; Q[T_f, T_\text{CMB}] \rangle$. 
However, this auto-correlation is biased by the large noise of $Q[T_f, T_\text{CMB}]$, which would have to be subtracted accurately.
We therefore propose and implement a new method to avoid this issue.
We Taylor-expand the lensed CMB map $T_\text{CMB} = T^0 + T^1+...$ in powers of $\kappa$,
and compute the quantity $8\langle Q[T_f, T^0] \; Q[T_f, T^1] \rangle$
\footnote{
Another way to evaluate the secondary bispectrum term
would be $\langle Q[T_f, T_\text{CMB}]Q[T_f, T_\text{CMB}] - Q[T_f, T_\text{CMB}']Q[T_f, T_\text{CMB}'] \rangle$ where $T_\text{CMB}'$
is constructed from the same unlensed CMB realization as $T_\text{CMB}$ but lensed by an
independent $\kappa$ realization. 
}.
This works because the quadratic estimators are by construction unbiased when applied to the pair $(T^0, T^1)$, to first order in lensing.
This greatly reduces the noise, and this is a cross-correlation so no noise subtraction is needed (no $N^0$, or higher order bias $N^i$).

3) For the \textit{trispectrum} term, we compute $\langle Q[T_f, T_f] \; Q[T_f, T_f] \rangle$, and subtract the Gaussian contribution (which is a part of $N^0$) analytically, as in \cite{2014ApJ...786...13V, 2014JCAP...03..024O}.

For the cross-correlation with tracers $C_L^{\kappa \delta_g}$, 
only the primary bispectrum is present, and without the combinatorial factor 2: $\langle Q[T_f, T_f] \; \delta_g \rangle$.
The secondary bispectrum and trispectrum terms only act as a source of noise on this cross-correlation, not bias.

\subsection*{Results}

\begin{figure}[h!!!!]
\centering
\includegraphics[width=1\columnwidth]{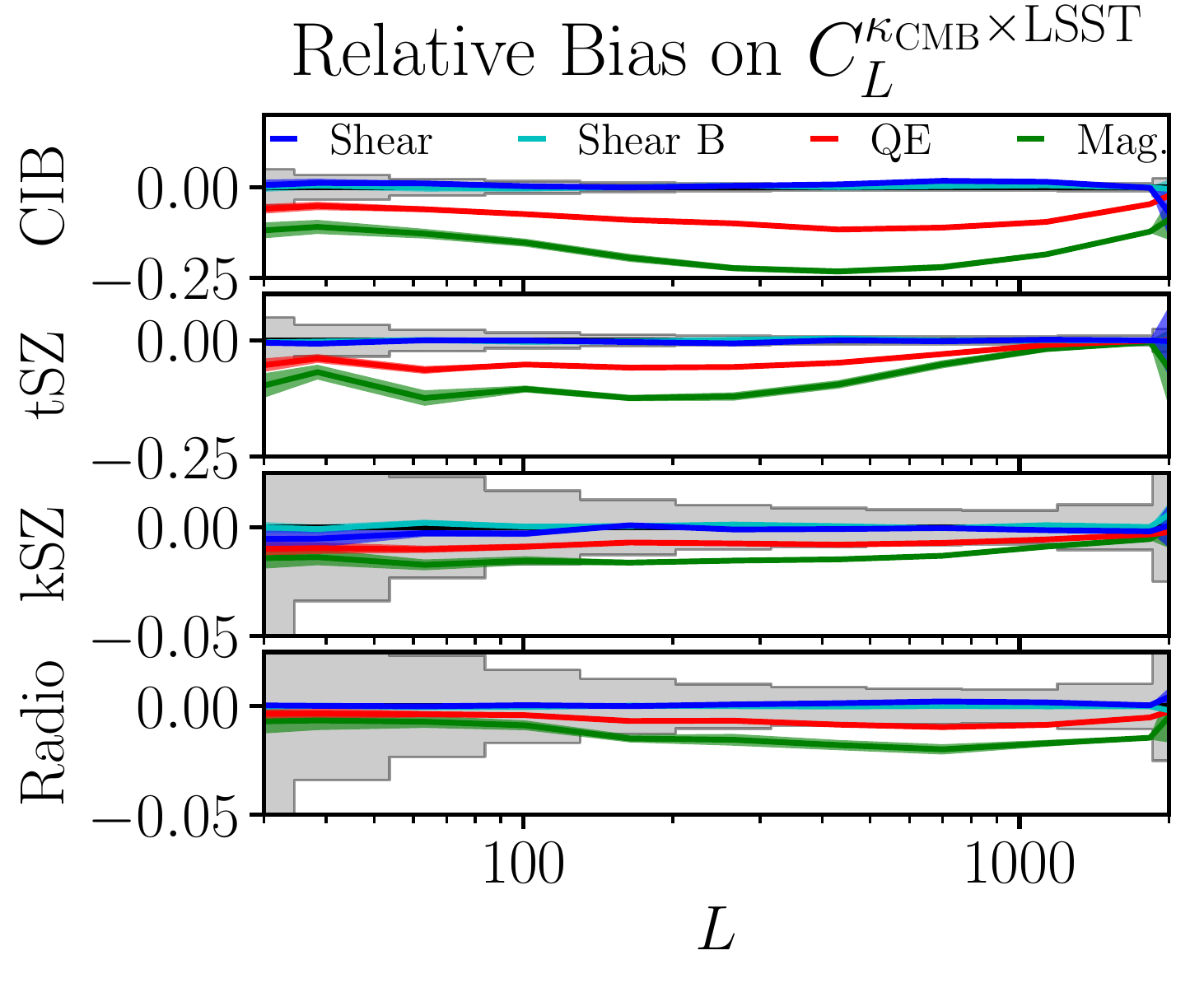}
\caption{
Relative bias to the cross-correlation between CMB lensing and the LSST gold galaxy sample, as a function of lensing multipole $L$,
when including temperature multipoles $\ell=30-3500$ at 148GHz. 
This foreground bias corresponds to the `primary bispectrum' term.
The grey boxes indicate bins of lensing multipoles with the corresponding statistical error bars for the standard quadratic estimator (lensing noise plus cosmic variance, identical in each panel).
The foreground biases are much larger than the statistical error bars for both the standard quadratic estimator and the magnification estimator, whereas they are barely measurable for the shear estimator.
}
\label{fig:summary_bias_cross_qsd_lmax3_5e3}
\end{figure}

The resulting foreground biases for the cross-correlation $C_L^{\kappa \delta_g}$ are shown in Fig.~\ref{fig:summary_bias_cross_qsd_lmax3_5e3}.
Despite the masking of point sources, the CIB, tSZ, kSZ and radio PS lead to very large and statistically significant biases 
for the QE and the magnification estimators.
Again, multi-frequency component separation may be used to null the tSZ bias, or reduce the CIB or radio PS biases. 
However,  reducing all these biases simultaneously typically causes a large noise increase.
Furthermore, multi-frequency analyses have no effect on the kSZ bias. 
These foreground biases are therefore a major concern for the standard QE.
On the other hand, no foreground bias is detected in the shear estimator. 
This is the main result of this letter: even when applied to a single-frequency temperature map, the shear estimator measures only the quadrupolar distortions from lensing, and is therefore immune to foregrounds. 
It is remarkable that this holds even for a single frequency map out to $\ell_{\text{max},T} = 3500$, where the temperature modes are foreground dominated.
Our QE tSZ bias in Fig.~\ref{fig:summary_bias_cross_qsd_lmax3_5e3} is smaller than in \cite{2018arXiv180208230M, 2018arXiv180205257B},
which can be explained by our scaling down of the tSZ map to match the power spectrum model of \cite{2013JCAP...07..025D},
our masking,
and the different redshift of our galaxy catalog.
Our CIB bias is slightly larger than found in \cite{2018arXiv180205257B}.

\begin{figure}[ht!!!!]
\centering
\includegraphics[width=0.9\columnwidth]{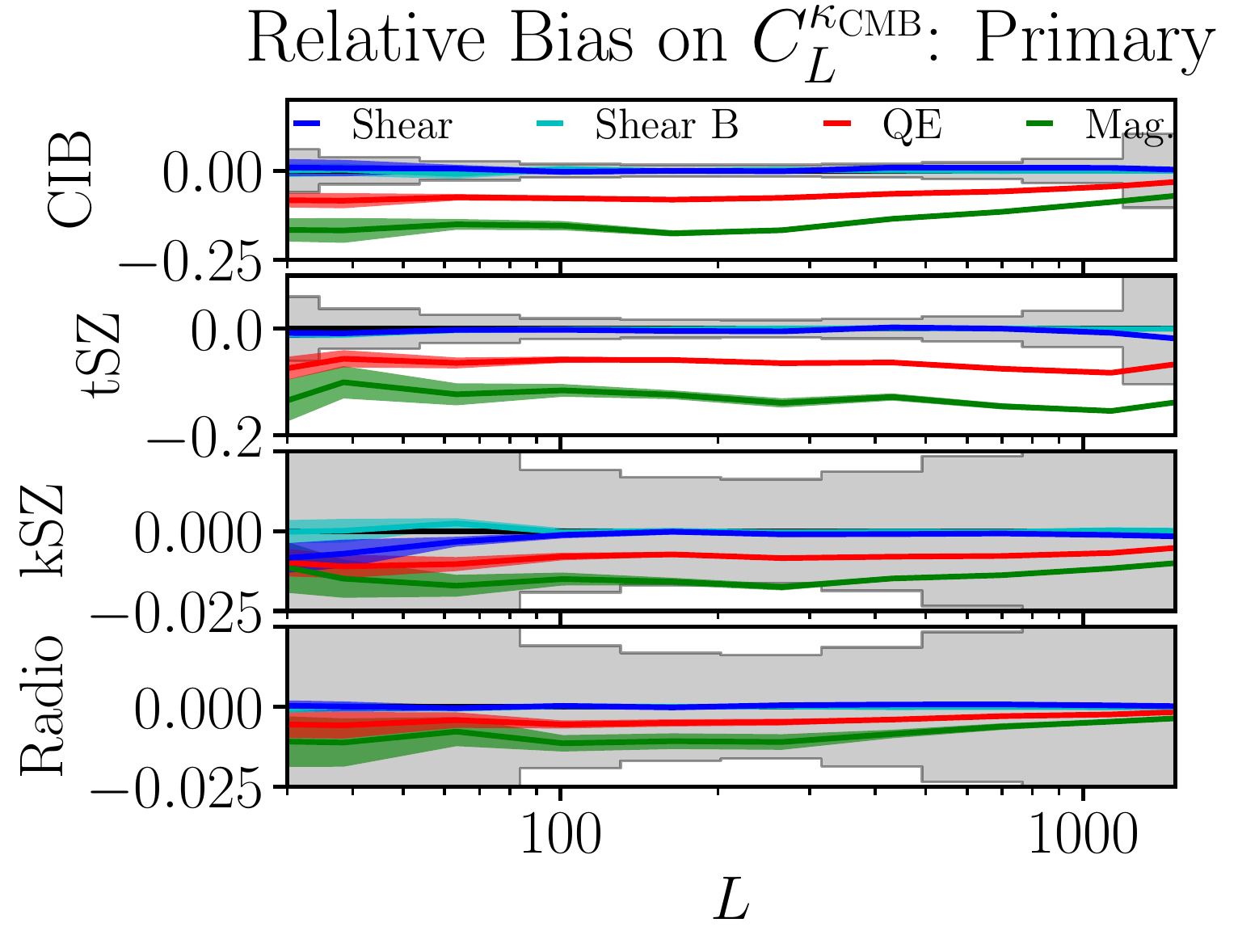}
\includegraphics[width=0.9\columnwidth]{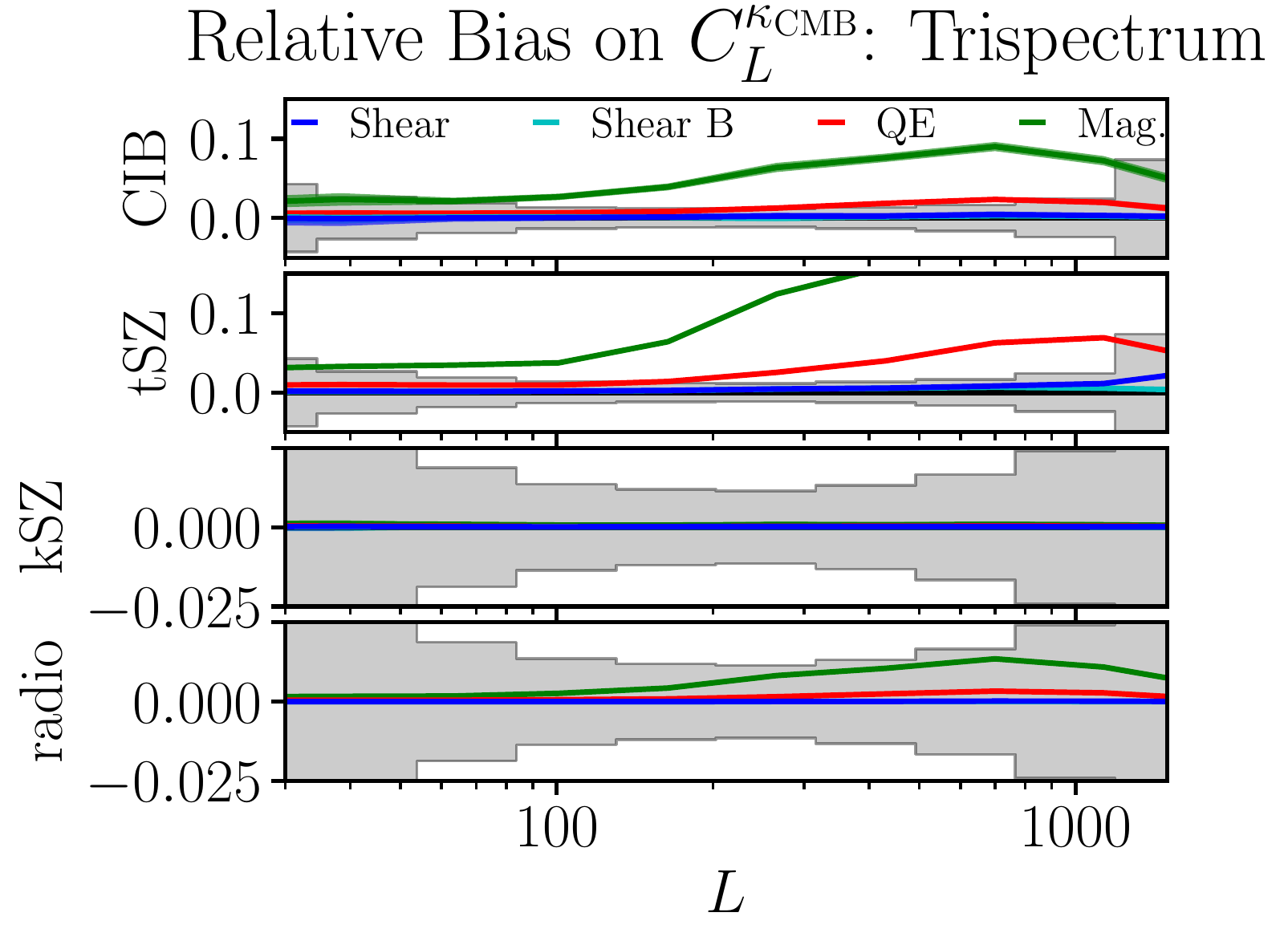}
\includegraphics[width=0.9\columnwidth]{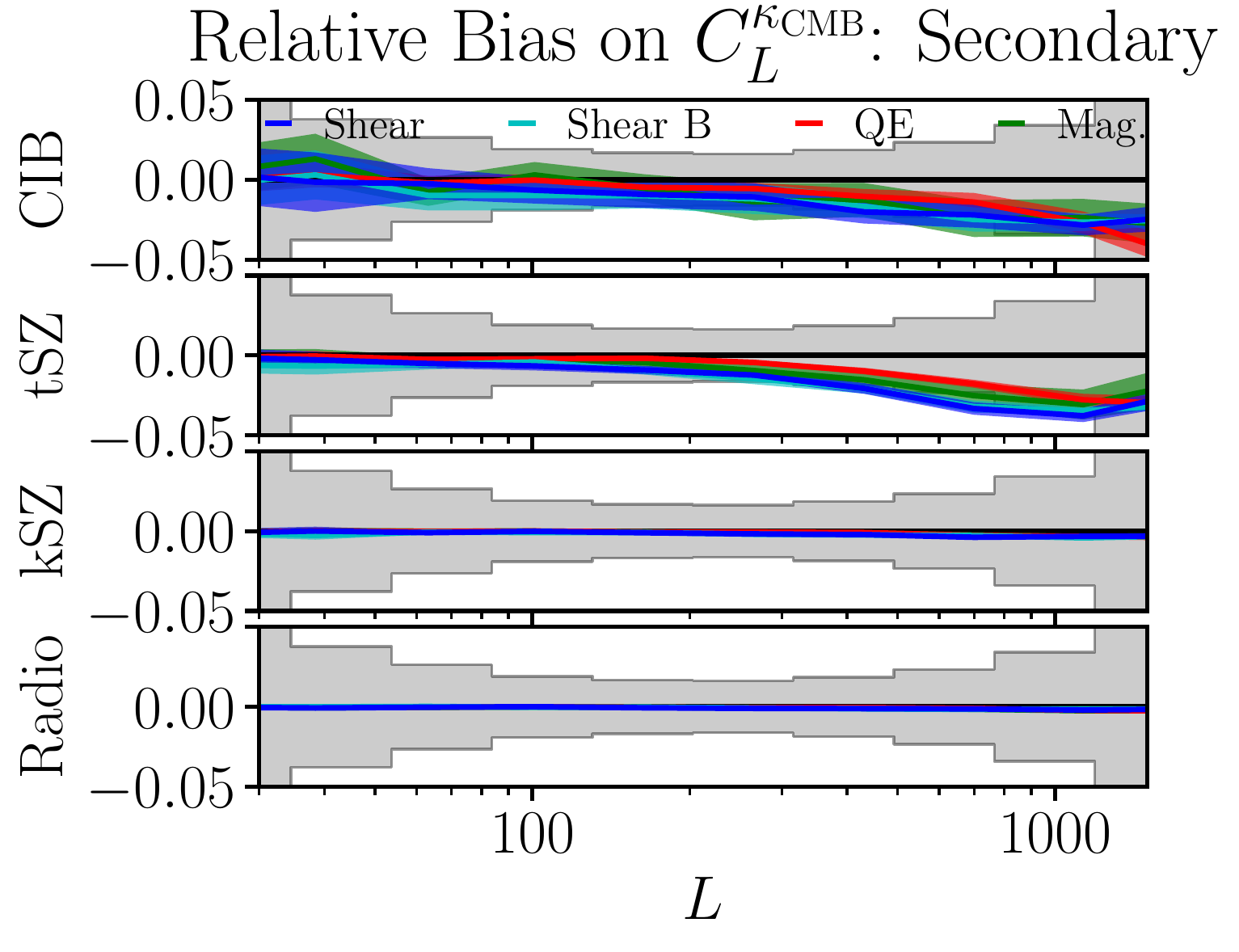}
\caption{
Relative foreground bias on the CMB lensing power spectrum, as a function of lensing multipole $L$, when including temperature multipoles $\ell=30-3500$ at 148GHz.
The grey boxes indicate bins of lensing multipoles with the corresponding statistical error bars for the standard quadratic estimator (lensing noise plus cosmic variance).
\textbf{Top:} primary bispectrum bias, dominant at low $L$.
\textbf{Middle:} trispectrum bias, dominant at high $L$.
\textbf{Bottom:} secondary bispectrum bias.\\
The dominant biases (primary and trispectrum) are much larger than the statistical error bars for the QE and magnification estimator, and are barely measurable for the shear estimator. 
The secondary bispectrum bias is smaller, and similar in size for all estimators. 
The secondary bispectrum bias is identical for the shear E and B estimators, making the difference of the two an unbiased lensing estimator.
}
\label{fig:summary_bias_prim_qsd_lmax3_5e3}
\end{figure}

For the lensing auto-spectrum $C_L^{\kappa\kappa}$, the primary, secondary and trispectrum biases discussed in the previous section are shown in Fig.~\ref{fig:summary_bias_prim_qsd_lmax3_5e3}.  
At low (resp. high) lensing multipoles, the primary (resp. trispectrum) bias dominates.
In both cases, a large bias is seen in the QE and magnification estimator, while the shear estimator is unbiased.
Our primary and trispectrum foreground biases are consistent with the results of \cite{2014ApJ...786...13V} for the CIB and tSZ, 
and slightly smaller than what found in \cite{2018PhRvD..97b3512F} for the kSZ, due to our rescaling of the kSZ map and the slightly different lensing weights.
We compute the secondary foreground bias separately.
This term is smaller than the primary and trispectrum term, but non-negligible for $L$ of a few hundred. 
Here, the shear estimator alone does not improve over the QE and magnification estimators.
This occurs because the shear secondary bias introduces a $\cos^2(2\theta)$, which makes it sensitive to the foreground monopole power.
However, the shear B-mode estimator has the same secondary bias and no response to lensing: 
subtracting it from the shear E-mode therefore cancels the secondary bias, at the cost of an increased noise.
Overall, the shear estimator dramatically reduces the foreground biases.
In the absence of any foreground cleaning, the shear estimator allows to increase the range of multipoles used in the lens reconstruction from $\ell_{\text{max},T} \approx 2500$ for the QE, to $\ell_{\text{max},T} \approx 3500$ for shear-only. 
Multi-frequency foreground cleaning may help increase the range of usable multipoles -- and thus the statistical power -- for all estimators.
The proposed shear B-mode subtraction may further improve the range for the shear E-mode estimator.
We leave a detailed optimization study to future work.

\section{Conclusion}

For current and upcoming CMB experiments such as AdvACT, SPT-3G and Simons Observatory,
CMB lensing reconstruction will rely heavily on temperature.
Foreground emission is known to contaminate temperature maps from which lensing is reconstructed, and therefore produce very significant biases, leading to wrong conclusions about cosmology if unaccounted for. 
Modeling and subtracting these bias terms is likely to be very challenging, due to the complex baryon physics involved in producing them. 
While some foregrounds can be nulled (tSZ) or reduced (CIB, radio PS) by a multi-frequency analysis, at the cost of a degradation in map noise, other foregrounds cannot (kSZ).

In this letter, we therefore explored a different approach,
by using the approximate isotropy of the extragalactic foreground 2d power spectra, 
and splitting the QE into optimal quadratic multipole estimators.

In the large-scale lens regime, they reduce to the isotropic magnification and anisotropic shear E-mode estimators of \cite{2008MNRAS.388.1819L,2012PhRvD..85d3016B, 2017arXiv170902227P}, and a new shear B-mode estimator.
The shear estimator enables a remarkable reduction of foreground biases, compared to the QE, even when applied to a single-frequency temperature map.
As a result, the shear estimator allows to increase the range of multipoles used in the lens reconstruction to $\ell_{\text{max},T} \approx 3500$, instead of $\ell_{\text{max},T} \approx 2500$ for the QE, while keeping foreground biases within the statistical uncertainty.
Overall, the signal-to-noise in shear with $\ell_{\text{max},T}=3500$ is very similar to that in QE with $\ell_{\text{max},T}=2500$.
The shear estimator thus provides a robust way of measuring lensing.
Component separation may allow the use of higher multipoles for all estimators.
On the other hand, the magnification estimator is highly sensitive to foregrounds, so comparing magnification and shear provides an excellent diagnostic for foreground contamination. 
The shear B-mode estimator constitutes an additional null test, and allows to further reduce foreground biases.
Quantifying the size of the higher order biases such as $N^{(1)}$ and $N^{(2)}$ for the shear and magnification estimators will be important.

Further optimization is possible, by combining different estimators with different $\ell_{\text{max},T}$.
For instance, a hybrid estimator magnification($\ell\leq 2000$) \& shear($\ell \leq3500$) improves the lensing signal-to-noise by $14\%$ compared to the standard QE($\ell\leq 2500$).

Better approximations to the optimal multipole estimators than the shear and magnification estimators may yield further improvements in signal-to-noise. A promising approach would be to replace the derivatives in Eq.~\eqref{eq:weights} by free functions of $\ell$ to be optimized.
Future CMB lensing data from CMB S4 should be polarization-dominated. The shear and magnification estimators can be generalized to polarization \cite{2017arXiv170902227P},
and may bring improvements there too.
This would have implications for precision delensing, in order to isolate primordial tensor modes.
Similar foreground biases occur in lens reconstruction from intensity mapping \cite{2018arXiv180205706S, 2018arXiv180304975F} (e.g., the `self-lensing bias' for CIB), and the shear estimator may allow to reduce them \cite{2018arXiv180205706S, 2018arXiv180304975F}.
Finally, the split into magnification and shear E and B-modes may also help detect residual Galactic foregrounds or beam ellipticity.
We leave the exploration of these promising avenues to future work.

\section*{Acknowledgments}

We thank Marcelo Alvarez, Anthony Challinor, Sandrine Codis, Simon Foreman, Colin Hill, Shirley Ho, Wayne Hu, Akito Kusaka, Antony Lewis, Heather Prince, Uro${\rm \check{s}}$ Seljak, David Spergel, Blake Sherwin, Alex van Engelen, Martin White and Hong-Ming Zhu for useful discussion. 
We thank the anonymous referees for very useful comments and suggestions, which greatly improved this paper. 
ES is supported by the Chamberlain fellowship at Lawrence Berkeley National Laboratory.
SF was in part supported by a Miller Fellowship at the University of California, Berkeley and by the Physics Division at Lawrence Berkeley National Laboratory.  
This work used resources of the National Energy Research Scientific Computing Center, a DOE Office of Science User Facility supported by the Office of Science of the U.S. Department of Energy under Contract No. DE-AC02-05CH11231.



\appendix

\onecolumngrid

\section{CMB lensing: review and link with magnification and shear E \& B modes}

The lensed temperature map $T$ is related to the unlensed map $T^0$ through $T(\n) = T^0(\n - \vd )$.
We then apply the usual 2d Helmholtz decomposition to the deflection field:
$\vd = \vec{\nabla}\phi + \vec{\nabla}\times (\omega \hat{e}_z)$.
The gradient term is known to produce convergence and E-mode shear, while the curl term produces B-mode shear and rotation, and vanishes in the Born approximation.
We define the usual convergence field $\kappa \equiv \frac{1}{2}\nabla^2 \phi$, and analogously for the curl term: $\kappa^\omega \equiv \frac{1}{2}\nabla^2 \omega$.

The deflection field breaks the statistical isotropy of the temperature map, and produces off-diagonal covariances:
\beq
\langle T_\vl T_{\vL-\vl} \rangle
=
f^\kappa_{\vl, \vL-\vl} \kappa_\vL
+
f^{\kappa^\omega}_{\vl, \vL-\vl} \kappa^\omega_\vL
+
\mathcal{O}\left( \kappa^2, \kappa^{\omega 2}, \kappa\kappa^\omega \right),
\label{eq:lensed_tt}
\eeq
with:
\beq
\left\{
\bal
&f^\kappa_{\vl_1, \vl_2}
\equiv
\left(
\frac{2 \vL}{L^2}
\right)
\cdot
\left[
\vl_1  C^0_{\ell_1}
+
\vl_2  C^0_{\ell_2}
\right]\\
&f^{\kappa^\omega}_{\vl_1, \vl_2}
\equiv
\left(
\frac{-2 \vL}{L^2}
\right)
\times
\left[
\vl_1  C^0_{\ell_1}
+
\vl_2  C^0_{\ell_2}
\right],\\
\eal
\right.
\eeq
and $\vL = \vl_1+\vl_2$.

These expressions take a more intuitive meaning in the large-scale lens regime, where we consider the effect of a large-scale lensing mode $\vL$ on small scale temperature multipoles $\vl_1 = \vL/2+\vl$ and $\vl_2=\vL/2-\vl$, with $L\ll \ell$:
\beq
\left\{
\bal
&f^\kappa_{\vl_1, \vl_2}
=
C^0_\ell
\left[
\underbrace{ \frac{d \ln \ell^2C^0_\ell}{d \ln \ell}}
_{\substack{\text{magnification}}}
+
\underbrace{ \cos 2\theta_{\vL, \vl} \frac{d \ln C^0_\ell}{d \ln \ell}}
_{\substack{\text{E-mode shear}}}
\right]
 +\mathcal{O}\left(\left(\frac{L}{\ell}\right)^2 \right)\\
&f^{\kappa^\omega}_{\vl_1, \vl_2}
=
- C^0_\ell
\left[
\underbrace{ \sin 2\theta_{\vL, \vl} \frac{d \ln C^0_\ell}{d \ln \ell}}
_{\substack{\text{B-mode shear}}}
\right]
 +\mathcal{O}\left(\left(\frac{L}{\ell}\right)^2 \right) . \\
\eal
\right.
\label{eq:response_large_lens}
\eeq 
In the expression above, we have Taylor-expanded $C^0_{|\vl \pm \vL/2|}$ around $|\vl|$. 
Since the unlensed power spectrum has oscillations on scales $\sim 200$, this Taylor expansion can only be accurate for $L/2 \lesssim 200$, i.e. $L \lesssim 400$. Outside this range, the estimators remain unbiased, but may become suboptimal.
As expected, the gradient term in the lensing deflection causes magnification and E-mode shear. The curl term in the lensing deflection causes a B-mode shear. The expected rotation due to the curl term does not appear at this order, because the unlensed power spectrum is isotropic.

\section{Splitting the QE: Multipole lensing estimators}

We start again from Eq.~\eqref{eq:lensed_tt}, and we ignore for now any potential curl term in the lensing deflection:
\begin{equation}
\langle T_{\vl + \frac{\vL}{2}} T_{\frac{\vL}{2} - \vl} \rangle
=
f_{\vl + \frac{\vL}{2} ,  \frac{\vL}{2} - \vl}^\kappa \;
\kappa_{\vL} 
+\mathcal{O}(\kappa^2).\\
\end{equation}
For a fixed $\vL$, the response function $f_{\vl + \frac{\vL}{2} ,  \frac{\vL}{2} - \vl}^\kappa$ is a function of $\vl$, or equivalently of $\left( \ell, \theta \right)$, where $\theta$ is the angle between $\vL$ and $\vl$.
We may therefore expand the $\theta$-dependence as a Fourier series:
\beq
f_{\vl + \frac{\vL}{2} ,  \frac{\vL}{2} - \vl}^\kappa
=
\sum_{m \in 2 \field{N}}
f_{L, \ell}^m
\cos \left( m \theta_{\vL, \vl} \right),
\text{     with   }
f_{L, \ell}^m
\equiv
\left\{
\bal
&\int \frac{d \theta}{\left( 2\pi \right)}
f_{\vl + \frac{\vL}{2} ,  \frac{\vL}{2} - \vl}^\kappa
\text{ if  } m=0
\\
&2
\int \frac{d \theta}{\left( 2\pi \right)}
f_{\vl + \frac{\vL}{2} ,  \frac{\vL}{2} - \vl}^\kappa
\cos \left( m \theta \right)
\text{ otherwise}
\\
\eal
\right.
.
\label{eq:multipole_expansion_lensing_response}
\eeq
This Fourier series only includes even modes $m\in 2\field{N}$, and cosine terms (no sine terms), because the function $f_{\vl + \frac{\vL}{2} ,  \frac{\vL}{2} - \vl}^\kappa$ is invariant under the transformations $\theta \rightarrow \theta + \pi$ and $\theta \rightarrow -\theta$, respectively.

In principle, the modes $f_{L, \ell}^m$ for all $m\in 2\field{N}$ may be non-zero, and we may estimate lensing using the information in the $m$-th mode. 
In what follows, we call $f_{L, \ell}^m$ the $m$-th multipole of the lensing response function $f^\kappa$,
and we derive the  minimum-variance unbiased quadratic estimator that reconstructs the convergence field from the $m$-th multipole only.
Such an estimator necessarily involves the angular average
%
$
\int \frac{d\theta}{(2\pi)} 
T_{\vl + \frac{\vL}{2}} T_{\frac{\vL}{2} - \vl} 
\cos(m\theta),
$
%
where $\theta$ is again the angle between $\vL$ and $\vl$.
Once normalized to be unbiased, this building block estimator becomes:
\beq
\hat{\kappa}_{\vL, \ell}^m
\equiv
\frac{\left(2 \text{ if } m>0\right)}
{f_{L, \ell}^m}
\int \frac{d\theta}{(2\pi)} 
T_{\vl + \frac{\vL}{2}} T_{\frac{\vL}{2} - \vl} 
\cos(m\theta).
\eeq
The minimum-variance estimator from the $m$-th multipole is then obtained by inverse-variance weighting. 
The variance is given by:
\beq
\text{Cov}
\left[
\hat{\kappa}_{\vL, l}^m,
\hat{\kappa}_{\vL ', l'}^{m}
\right]
= 
(2\pi)^2 \delta^D_{\vL+\vL '} \;
(2\pi)^2 \frac{\delta^D_{\ell -\ell'}}{2} \;
\underbrace{
\frac{2}{\ell}
\frac{\left(4 \text{ if }m>0\right)}{ \left(f_{L, \ell}^m \right)^2}
\int \frac{d\theta}{(2\pi)} 
C^\text{total}_{|\vl + \frac{\vL}{2}|} C^\text{total}_{|\frac{\vL}{2} - \vl |}
\cos^2(m\theta)}
_{\equiv \sigma_{L, l}^{m\; 2}}
\label{eq:cov_kLlm}
\eeq
Hence the minimum-variance unbiased estimator from the $m$-th moment only can be written as:
\beq
\hat{\kappa}_\vL^m
\equiv
\frac{
\int \frac{d\ell}{(2\pi)}
\hat{\kappa}_{\vL, l}^m
/ \sigma_{L, l}^{m\; 2}
}
{
\int \frac{d\ell}{(2\pi)}
1
/ \sigma_{L, l}^{m\; 2}
}
,
\text{                        with                     }
N_\vL^{m}
=
\left[
\int \frac{d\ell}{(2\pi)}
1
/ \sigma_{L, l}^{m\; 2}
\right]^{-1}
.
\eeq
Or more explicitly:
\beq
\hat{\kappa}_\vL^m
=
N_\vL^{m} \;
\int \frac{d\ell}{(2\pi)}
\ell
\frac{f_{L, \ell}^m}
{
\left(
2
\int \frac{d\theta '}{(2\pi)} 
C^\text{total}_{|\vl ' + \frac{\vL}{2}|} C^\text{total}_{|\frac{\vL}{2} - \vl ' |}
\cos^2(m\theta ')
\right)
}
\frac{1}{\left(2 \text{  if  }m>0\right)}
\int \frac{d\theta}{(2\pi)} 
T_{\vl + \frac{\vL}{2}} T_{\frac{\vL}{2} - \vl} 
\cos(m\theta)
,
\label{eq:optimal_quadratic_multipole_estimator}
\eeq
with:
\beq
N_\vL^{m}
=
\left[
\int \frac{d\ell}{(2\pi)}
\ell
\frac{
\left(f_{L, \ell}^m \right)^2 /  \left(4 \text{  if }m>0\right)
}
{
\left(
2
\int \frac{d\theta '}{(2\pi)} 
C^\text{total}_{|\vl ' + \frac{\vL}{2}|} C^\text{total}_{|\frac{\vL}{2} - \vl ' |}
\cos^2(m\theta ')
\right)
}
\right]^{-1}
.
\eeq
\begin{figure}[h!!!]
\centering
\includegraphics[width=0.45\columnwidth]{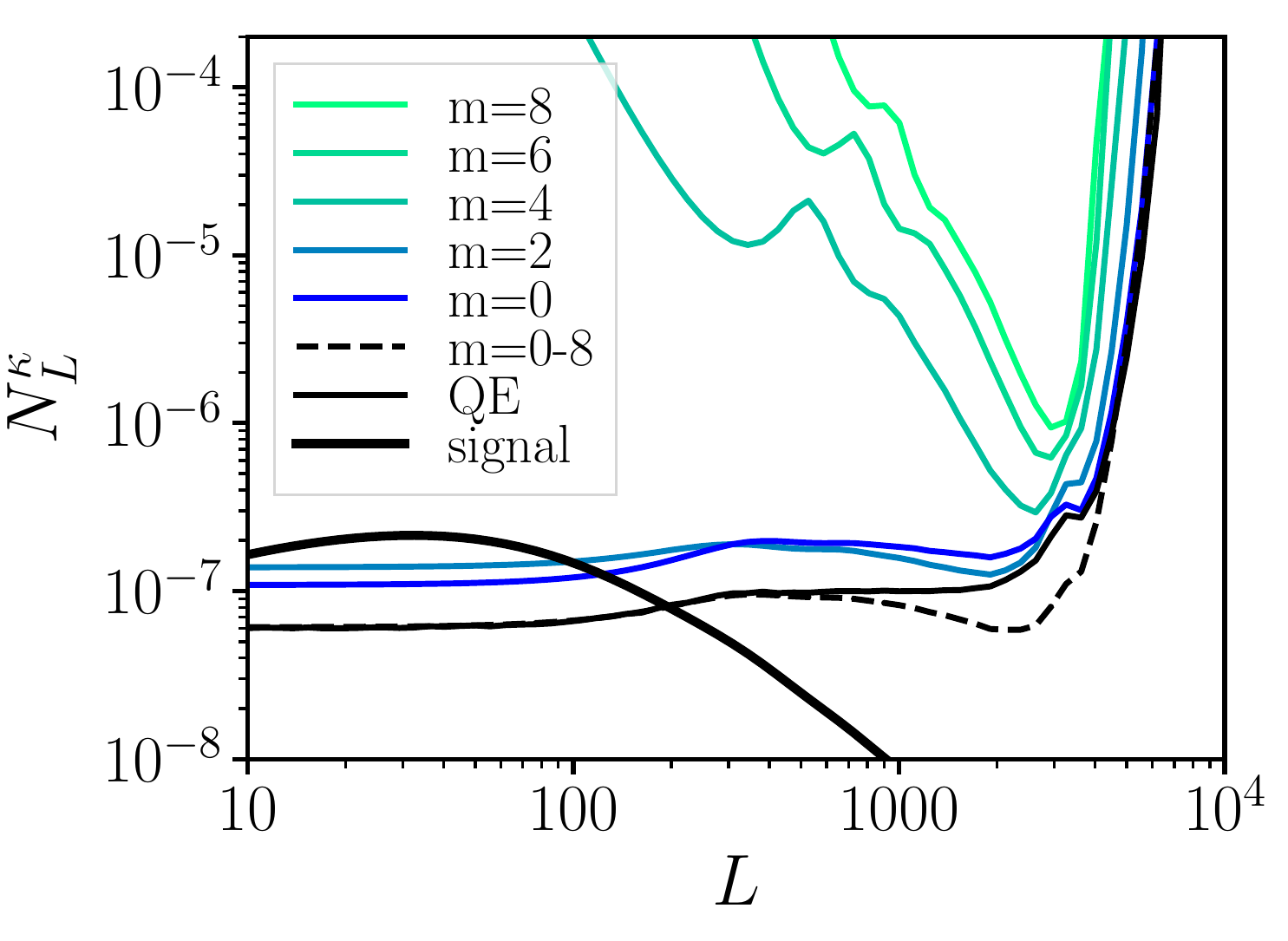}
\includegraphics[width=0.45\columnwidth]{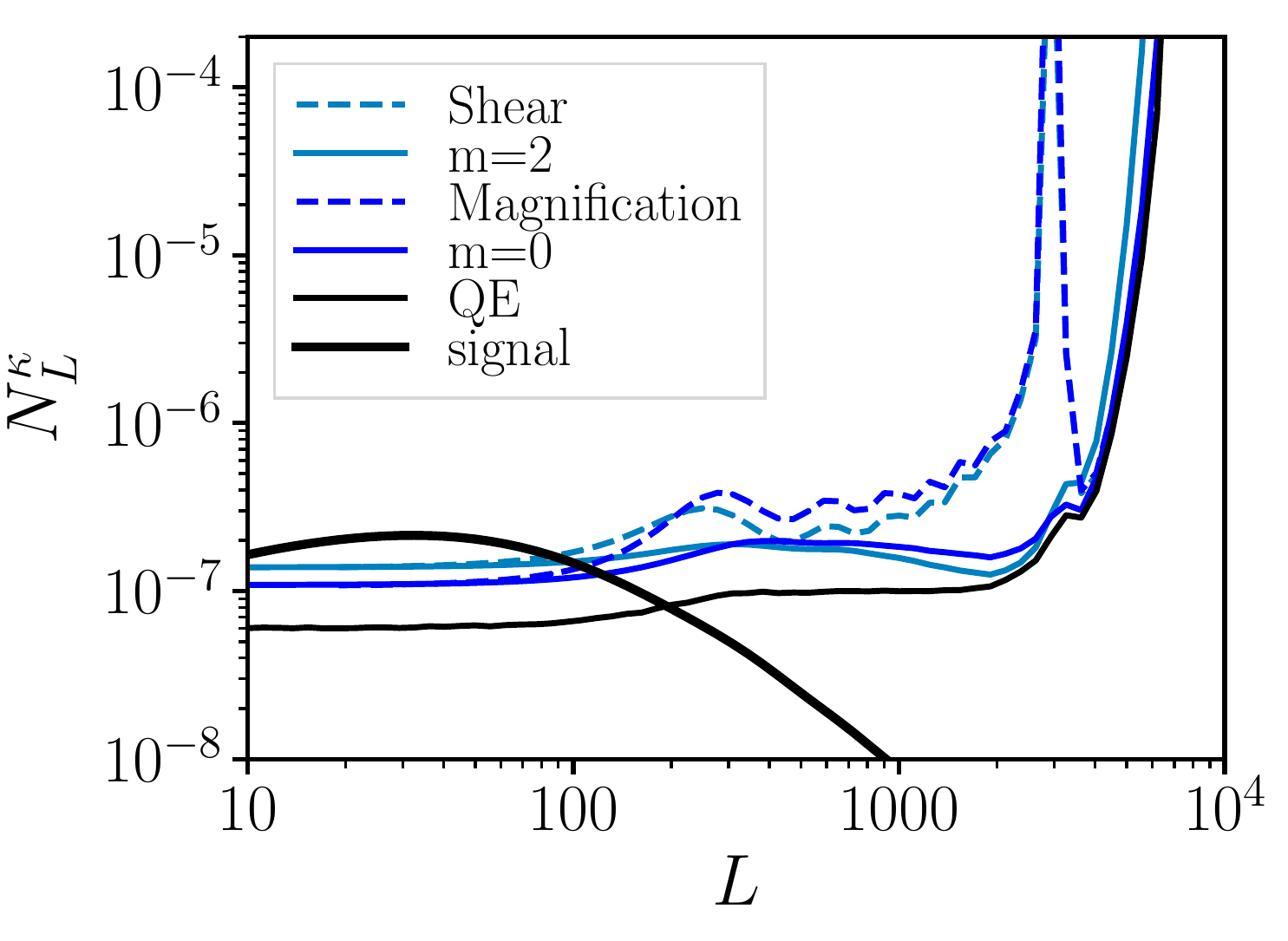}
\caption{
\textbf{Left:}
Reconstruction noise for the various multipole estimators. The monopole ($m=0$) and quadrupole ($m=2$) contain most of the lensing information.
The black dashed line shows the expected noise when combining the multipole estimators from $m=0$ to $m=8$, ignoring their covariances as appropriate in the squeezed limit. Indeed, for lensing multipoles $L \leq 300$, the multipole estimators are roughly uncorrelated, and recover the statistical uncertainty of the QE.
At higher lensing multipoles, the estimators are no longer uncorrelated, and the dashed line underestimates the statistical uncertainty.\\
\textbf{Right:}
Comparison of the lensing noise for the optimal monopole and quadrupole estimators (solid lines), and the suboptimal but faster to evaluate magnification and shear estimators (dashed lines).
In the large-scale lens regime, the shear and dilation estimators are equivalent to their optimal counterparts. 
This is no longer the case at higher lensing multipoles, where their noise spectra show a distinctive spike at $L\simeq 3000$.
This is due to the first order Taylor expansion in $L/\ell$ breaking down, and accidentally nulling the response of the estimators to lensing at this specific lensing multipole.
}
\label{fig:multipole_lensing_noises}
\end{figure}

Furthermore, the lensing multipole estimators are uncorrelated in the squeezed limit $L\ll \ell$, as can be shown by generalizing Eq.~\eqref{eq:cov_kLlm} to $m'\neq m$. Indeed, in this limit, the angular integral becomes
$\int \frac{d\theta}{(2\pi)} (C^{\text{total}}_\ell)^2 \cos(m\theta) \cos(m' \theta) \propto \delta^K_{m, m'} $.
We have therefore built a family of lensing estimators.
As shown in Fig.~\ref{fig:multipole_lensing_noises}, the monopole and quadrupole estimators contain most of the lensing signal-to-noise.
In what follows, we therefore focus on these two estimators.

Splitting the QE into this family of multipole estimators may be useful,
because some multipoles may be more affected by foregrounds than others.
In particular, we expect that the monopole estimator $m=0$ will be most sensitive to foregrounds, while higher multipole estimators should be more robust.
We verify this hypothesis in this paper, and propose using only the quadrupole lensing estimator, in order to avoid foreground biases.
We could also derive the optimal estimator from all lensing multipoles but the monopole.
However, since monopole and quadrupole contain most of the signal-to-noise, that estimator would not improve much over the quadrupole-only estimator.

While we focused on the gradient term of the lensing distortion, a similar analysis can be performed for the curl term of the lensing distortion. In that case, the Fourier series contains only $\sin$ terms instead of only $\cos$ terms, starting at $m=2$.
In the single lens plane approximation and in the Born approximation, the  multipole estimators of the curl term constitute useful null tests for the corresponding multipole estimators of the gradient term.

\section{Large-scale lens limit: recovering the shear and magnification estimators}

The quadratic multipole estimators of Eq.~\eqref{eq:optimal_quadratic_multipole_estimator} are optimal, but they cannot easily be evaluated with fast Fourier transform, because they are not easily recast as a sum of products or convolutions.
For this reason, we instead evaluate the following approximate estimators.

In the large-scale lens regime, i.e. when $L \ll \ell$, several elements of Eq.~\eqref{eq:optimal_quadratic_multipole_estimator} simplify.
First, the lensing response reduces to Eq.~\ref{eq:response_large_lens}, which only has a monopole and quadrupole.
In other words, the multipole expansion Eq.~\eqref{eq:multipole_expansion_lensing_response} of the lensing response simplifies:
\beq
\left\{
\bal
&f_{L, \ell}^0 = C^0_\ell \; \frac{d \ln \ell^2C^0_\ell}{d \ln \ell} + \mathcal{O}\left( \left(\frac{L}{\ell}\right)^2 \right)\\
&f_{L, \ell}^2 = C^0_\ell \;  \frac{d \ln C^0_\ell}{d \ln \ell} + \mathcal{O}\left( \left(\frac{L}{\ell}\right)^2 \right)\\
&f_{L, \ell}^m = 0 + \mathcal{O}\left( \left(\frac{L}{\ell}\right)^2 \right)  \text{ for }m \geq 4.\\
\eal
\right.
\eeq

The noise weighting in the optimal multipole estimators also simplifies
\footnote{Here, it may have seemed more natural to use $2 C^\text{total}_{|\vl |} C^\text{total}_{|\vL - \vl |}$ instead of $2 (C^{\text{total}}_\ell)^2$, so that each leg of the quadratic estimator is noise weighted. However, we checked that our choice actually leads to a slightly higher signal-to-noise.}
\beq
2
\int \frac{d\theta '}{(2\pi)} 
C^\text{total}_{|\vl + \frac{\vL}{2}|} C^\text{total}_{|\frac{\vL}{2} - \vl |}
\cos^2(m\theta)
\times
\left(2 \text{  if  }m>0\right)
\rightarrow
2 \; (C^{\text{total}}_\ell)^2
+ 
\mathcal{O}\left( \left(\frac{L}{\ell}\right)^2 \right).
\label{eq:noise_weighting_large_lens}
\eeq

The optimal estimators of Eq.~\eqref{eq:optimal_quadratic_multipole_estimator} therefore take a simpler form.
Since the denominator is only a normalization, we focus on the numerator.
Eq.~\eqref{eq:response_large_lens} and Eq.~\eqref{eq:noise_weighting_large_lens} are valid up to a $\mathcal{O}\left( \left( \frac{L}{\ell} \right)^2 \right)$ correction.
Combining them, we thus obtain the following approximation, valid to the same order in the large-scale lens regime:
\beq
\hat{\kappa}_\vL^m
\propto
\int
\frac{d^2\vl}{(2\pi)^2} \;
g_{\vL,\vl}^m \;
T_\vl
T_{\vL-\vl}
,
\text{ with }
\left\{
\bal
&g_{\vL,\vl}^0 \equiv \frac{C^0_\ell}{(C^{\text{total}}_\ell)^2} \; \frac{d \ln \ell^2C^0_\ell}{d \ln \ell} \\
&g_{\vL,\vl}^2 \equiv \frac{C^0_\ell}{(C^{\text{total}}_\ell)^2} \; \frac{d \ln C^0_\ell}{d \ln \ell} \cos(2\theta_{\vL, \vl})\\
\eal
\right.
.
\label{eq:numerator_shear_magnification}
\eeq
In Eq.~\eqref{eq:numerator_shear_magnification}, we have also replaced $\langle T_{\vl + \frac{\vL}{2}} T_{\frac{\vL}{2} - \vl}  \rangle$ with $\langle T_\vl
T_{\vL-\vl} \rangle$, to turn the integral into a convolution and allow a fast evaluation with FFT. 
This replacement would naively introduce a correction of order $\mathcal{O}\left( \frac{L}{\ell} \right)$:
\beq
\langle T_{\vl } T_{\vL - \vl} \rangle
\simeq
C^0_\ell
\left[
\underbrace{\kappa_\vL}
_{\substack{\text{isotropic} \\ \text{magnification}}}
\frac{\partial \ln \ell^2C^0_\ell}{\partial \ln \ell}
+
\underbrace{\kappa_\vL \cos 2\theta_{\vL, \vl} }
_{\substack{\text{anisotropic} \\ \text{shear}}}
\frac{\partial \ln C^0_\ell}{\partial \ln \ell}
+
\underbrace{\kappa_\vL \cos \theta_{\vL, \vl}}
_{\substack{\text{additional term}\\ \text{which averages to zero}}}
\frac{2L}{\ell} 
\frac{\partial \ln C^0_\ell}{\partial \ln \ell}
+
\mathcal{O} \left( \frac{L}{\ell} \right)^2
\right].
\eeq
However, this cosine term cancels inside the integral.
As a result, the approximate estimators of Eq.~\eqref{eq:numerator_shear_magnification} are still correct up to a  $\mathcal{O}\left( \left( \frac{L}{\ell} \right)^2 \right)$ correction.
We then normalize these approximate estimators so that they are exactly unbiased (to first order in lensing, just like the QE)
both in and out of the large-scale lens regime:
\beq
\hat{\kappa}_\vL^m
=
\frac{
\int
\frac{d^2\vl}{(2\pi)^2} \;
g_{\vL,\vl}^m \;
T_\vl
T_{\vL-\vl}
}
{
\frac{2\vL}{L^2} 
\cdot
\int
\frac{d^2\vl}{(2\pi)^2} \;
g_{\vL, \vl}^m
\left[
\vl  C^0_{\ell}
+
\left( \vL - \vl \right)  C^0_{|\vL-\vl |}
\right]
}
.
\label{eq:normalized_shear_magnification}
\eeq
We thus recover the magnification ($m=0$) and shear ($m=2$) estimators of \cite{2012PhRvD..85d3016B,2017arXiv170902227P}, although corrected for their multiplicative bias.
These estimators can be evaluated efficiently using fast Fourier transforms, since they are sums of products and convolutions.
In what follows, we assess the sensitivity of these estimators to residual foregrounds in the CMB map used for lensing reconstruction.

The noise power spectrum for the magnification and shear estimator can be computed as:
\beq
N_L^{\kappa^m}
=
\frac{
\int
\frac{d^2\vl}{(2\pi)^2} \;
g_{\vL, \vl}^m \left( g_{\vL, \vl}^m + g_{\vL, \vL-\vl}^m \right) \;
C^\text{total}_\ell C^\text{total}_{\vL-\vl}
}
{
\left\{
\frac{2\vL}{L^2} 
\cdot
\int
\frac{d^2\vl}{(2\pi)^2} \;
g_{\vL, \vl}^m
\left[
\vl  C^0_{\ell}
+
\left( \vL - \vl \right)  C^0_{|\vL-\vl |}
\right]
\right\}^2
}
.
\label{eq:noise_shear_magnification}
\eeq
This expression can be generalized easily to give the cross-spectrum between the noise of the shear and magnification estimators.
In Fig.~\ref{fig:multipole_lensing_noises}, we show that the noise power spectrum of the magnification and shear estimators match those of the monopole and quadrupole estimators respectively, for lensing multipoles $L\lesssim 100$, where the large-scale lens limit is valid. 
However, the magnification and shear estimators are suboptimal compared to the $m=0$ and $m=2$ multipole estimators for $L\gtrsim 100$, and show a distinctive spike in noise power spectrum. 
This occurs because the approximate lensing response $g^m_{\vL, \vl}$ is only accurate in the large-scale lens regime, i.e. at low lensing multipoles.
For higher lensing multipoles, the approximate lensing response has the wrong sign, going through a point where the approximate estimators have zero response to lensing, leading to an infinite noise.

Perhaps more surprisingly, Fig.~\ref{fig:multipole_lensing_noises} also shows that for high lensing multipoles $L \gtrsim 4000$, the noise power spectra of the QE, shear and magnification estimators become identical.
This can be understood as follows: to form a triangle configuration at such high lensing multipoles $L \gtrsim \ell_\text{max T}$, the temperature multipole themselves have to be $\sim \ell_\text{max T}$.
In this regime, $| \vl_1| \sim |\vl_2|$, so $\cos(2\theta_{\vL, \vl}) \sim 1$.
Furthermore, the unlensed/lensed CMB power spectrum at $\ell\sim \ell_\text{max T}$ is close to a power law ($\propto \ell^{-4}$).
Thus, the logarithmic derivatives of the CMB power spectrum (lensed or unlensed) that appear in the shear and magnification weights become a simple number, and have no effect on the weighting.
As a result, the three lensing estimators become identical in this limit.

Furthermore, a shear B-mode estimator can be obtained by replacing the $\cos$ term by a $\sin$ in the numerators of Eq.~\eqref{eq:numerator_shear_magnification}, \eqref{eq:normalized_shear_magnification} and \eqref{eq:noise_shear_magnification}.
As we show in this paper, this estimator has zero response to lensing and provides a useful null test to the shear E-mode estimator.
In particular, it allows to subtract any residual secondary foreground bias.

\begin{figure}[h!!!]
\centering
\includegraphics[width=0.4\columnwidth]{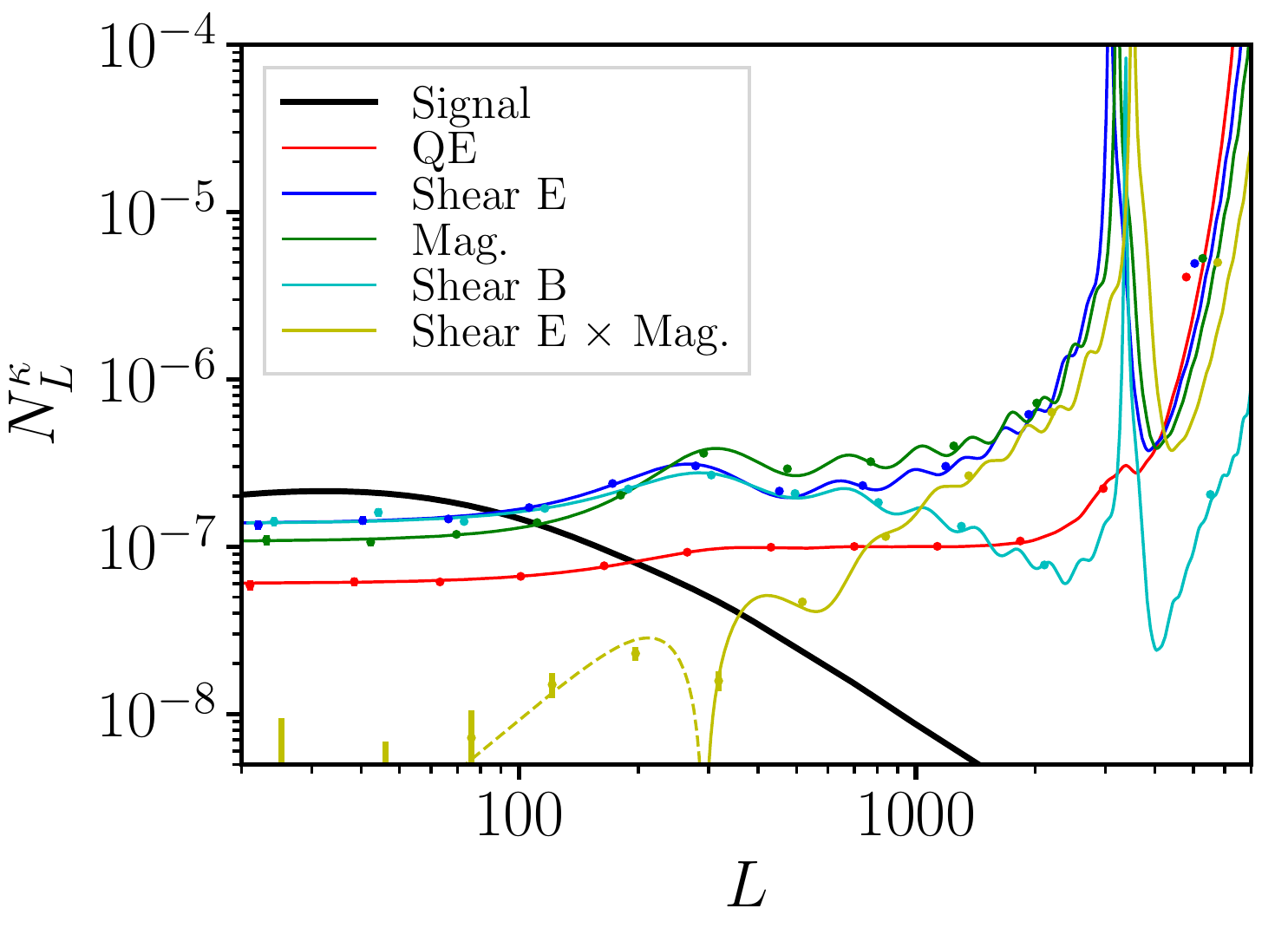}
\includegraphics[width=0.4\columnwidth]{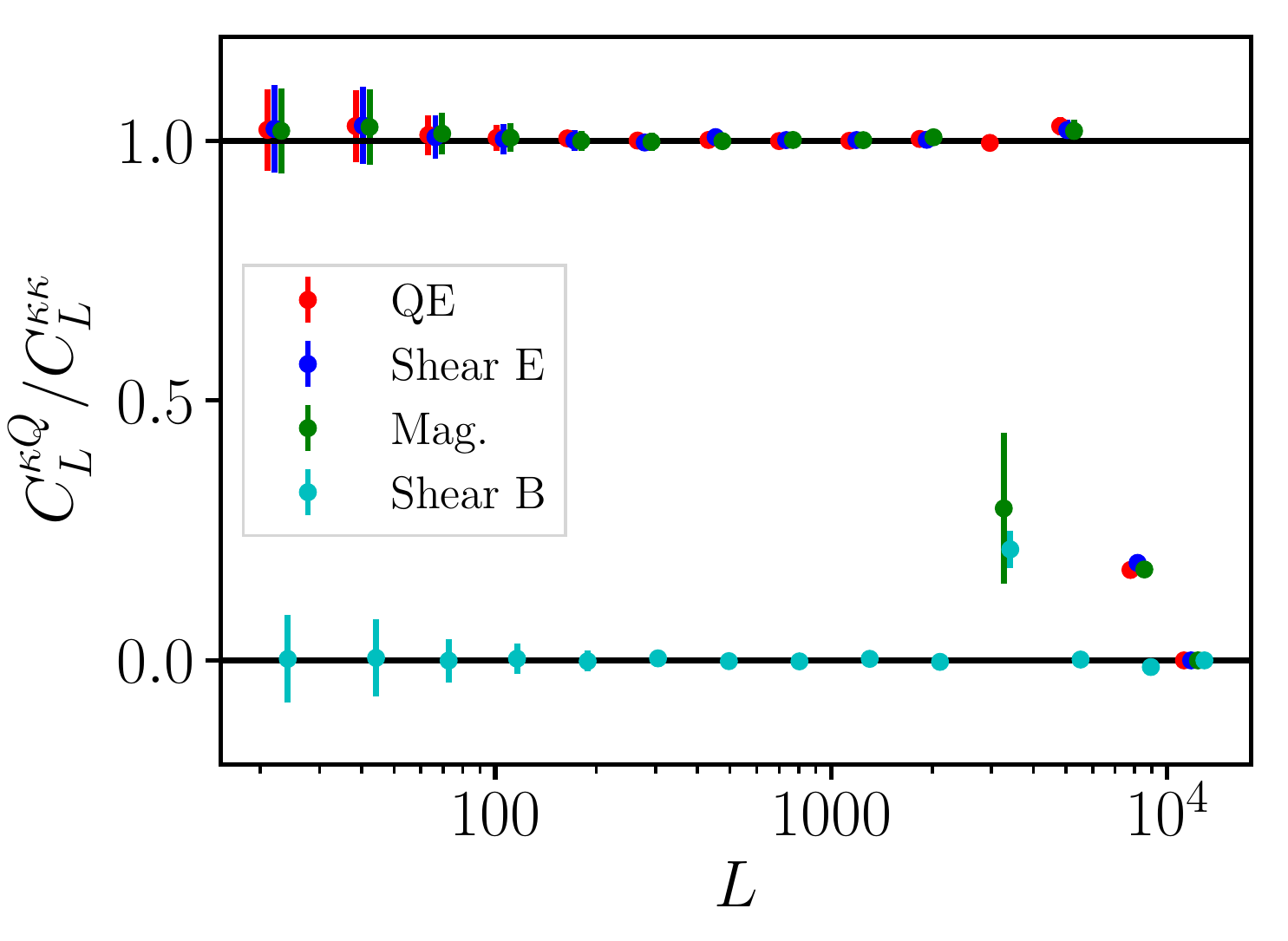}
\caption{
Validation of the pipeline for generating maps, lensing maps, and reconstructing lensing with the QE, shear E, shear B and magnification estimators.
We evaluate these estimators on 81 mock maps, square and periodic, of 400 deg$^2$ each.\\
\textbf{Left panel:} The theory expression for the reconstruction noises (curves) match the measured power spectrum of the various lensing estimators, when applied to a mock map with no lensing (Gaussian random field with power spectrum equal to the total power spectrum).\\
\textbf{Right panel:} The QE, shear E and magnification estimators have unit response to lensing, and the shear B estimator has zero response to lensing, as expected. The deviation at $L\simeq 3000$ corresponds to the noise spike in the shear and magnification estimators, where they effectively both have zero response to lensing. We expect this not to happen for the multipole estimators.
}
\label{fig:n0_qsd}
\end{figure}

In Fig.~\ref{fig:n0_qsd}, we show that these analytical expressions for the noise power spectrum match the measured power spectrum of the estimator,
when applied to mock Gaussian CMB maps with power spectrum equal to the total power spectrum (lensed CMB + detector noise + foregrounds).
We also show that these estimators are indeed unbiased, when applied to a Gaussian periodic unlensed CMB map, lensed by a Gaussian periodic lensing map.

\section{Foreground spectra}

\begin{figure}[h!]
\centering
\includegraphics[width=0.4\columnwidth]{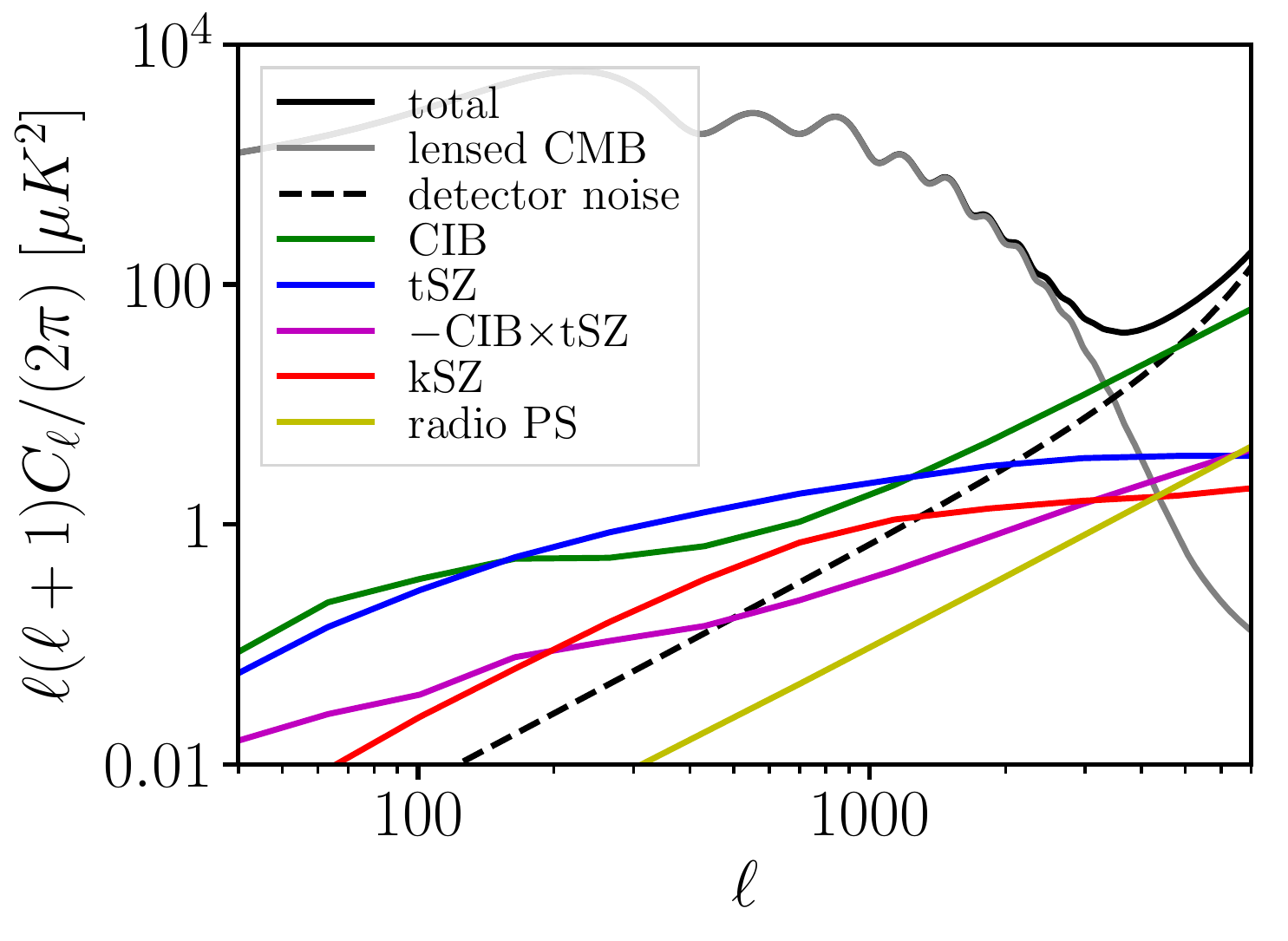}
\caption{
Power spectra of foregrounds used in this work. We fixed the normalization of the Sehgal maps \cite{2010ApJ...709..920S} to match the power spectrum model from \cite{2013JCAP...07..025D}. The Sehgal maps were subsequently masked for point sources above 5mJy, producing the solid curves shown in this figure.
At the power spectrum level, the effect of masking is most important for the radio point sources.
}
\label{fig:sehgal_dunkley_power_after_masking}
\end{figure}

We compute the power spectra of the various foreground maps from \cite{2010ApJ...709..920S}, before masking, and multiply them by factors of order unity (0.38 for CIB, 0.7 for tSZ, 0.82 for kSZ, 1.1 for radio PS) to match the spectra from \cite{2013JCAP...07..025D}.
After masking, the resulting power spectra are shown in Fig.~\ref{fig:sehgal_dunkley_power_after_masking}, and compared to the spectra from \cite{2013JCAP...07..025D}.
At the power spectrum level, the effect of masking is most spectacular for the radio PS.
However, while masking may not change the foreground power much, it may have a larger effect on the foreground bispectrum and trispectrum, thus reducing the foreground biases to CMB lensing.

\section{Galaxy catalog}

\begin{figure}[h!]
\centering
\includegraphics[width=0.4\columnwidth]{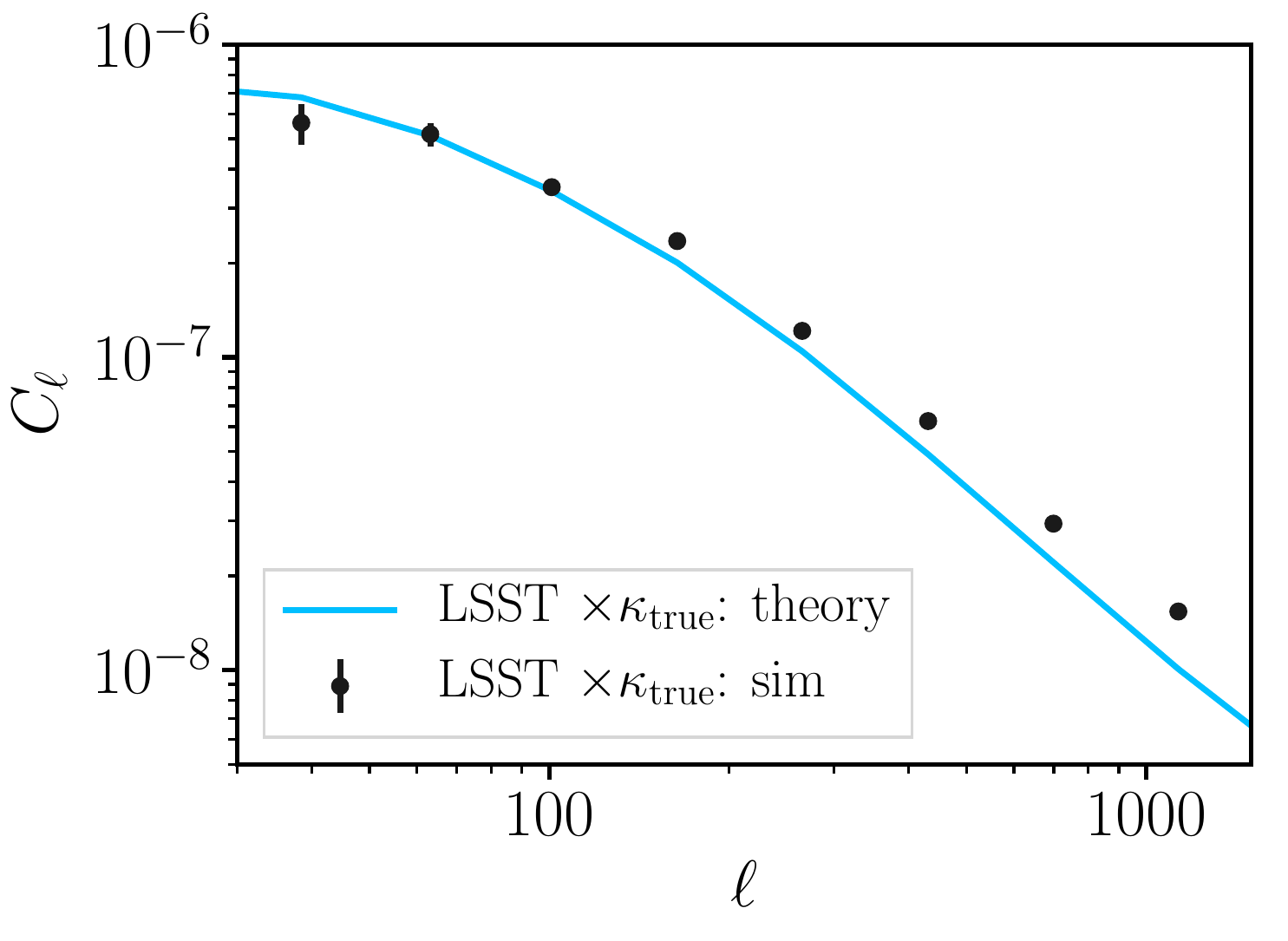}
\caption{
Cross correlation between our mock LSST gold sample, constructed by reweighting the halos in the catalog from \cite{2010ApJ...709..920S}, and the CMB lensing convergence in the same simulations.  The solid line shows the theory expectation for the actual LSST gold sample \cite{2009arXiv0912.0201L}. The rough agreement implies that the bias of the reweighted sample is close to the one of LSST sources, which is sufficient to estimate the foreground biases to LSST$\times \kappa_\text{CMB}$.}
\label{fig:check_lsstgold_bias}
\end{figure}

To construct a mock LSST gold sample, we re-weight the halos in the catalog from \cite{2010ApJ...709..920S} to match the redshift distribution of the LSST gold sample, with $i$-band magnitude $i < 25.3$ \cite{2009arXiv0912.0201L}:
\beq
\frac{dn}{dz} \propto \frac{1}{2 z_0} \left(\frac{z}{z_0}\right)^2 e^{-z/z_0}  \text{       , with   } z_0=0.24.
\eeq
The expected galaxy bias for the LSST sample is $b(z) = 1 + 0.84 z$ \cite{2009arXiv0912.0201L},
and Fig.~\ref{fig:check_lsstgold_bias} shows that our reweighted mock catalog has approximately the same bias.

\section{Foreground biases to the lensing amplitude}

\begin{figure}[h!!!]
\centering
\includegraphics[width=0.45\columnwidth]{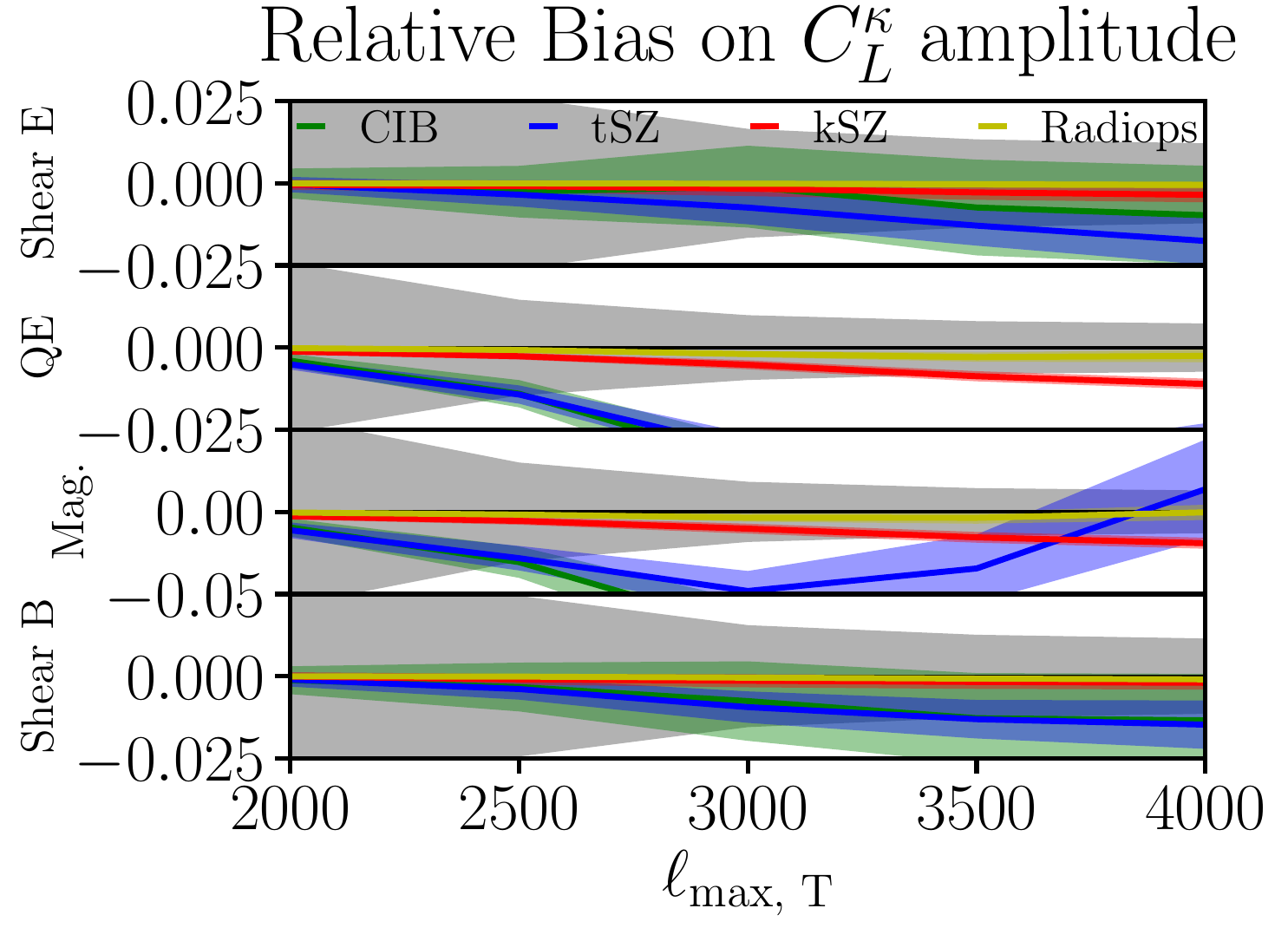}
\includegraphics[width=0.45\columnwidth]{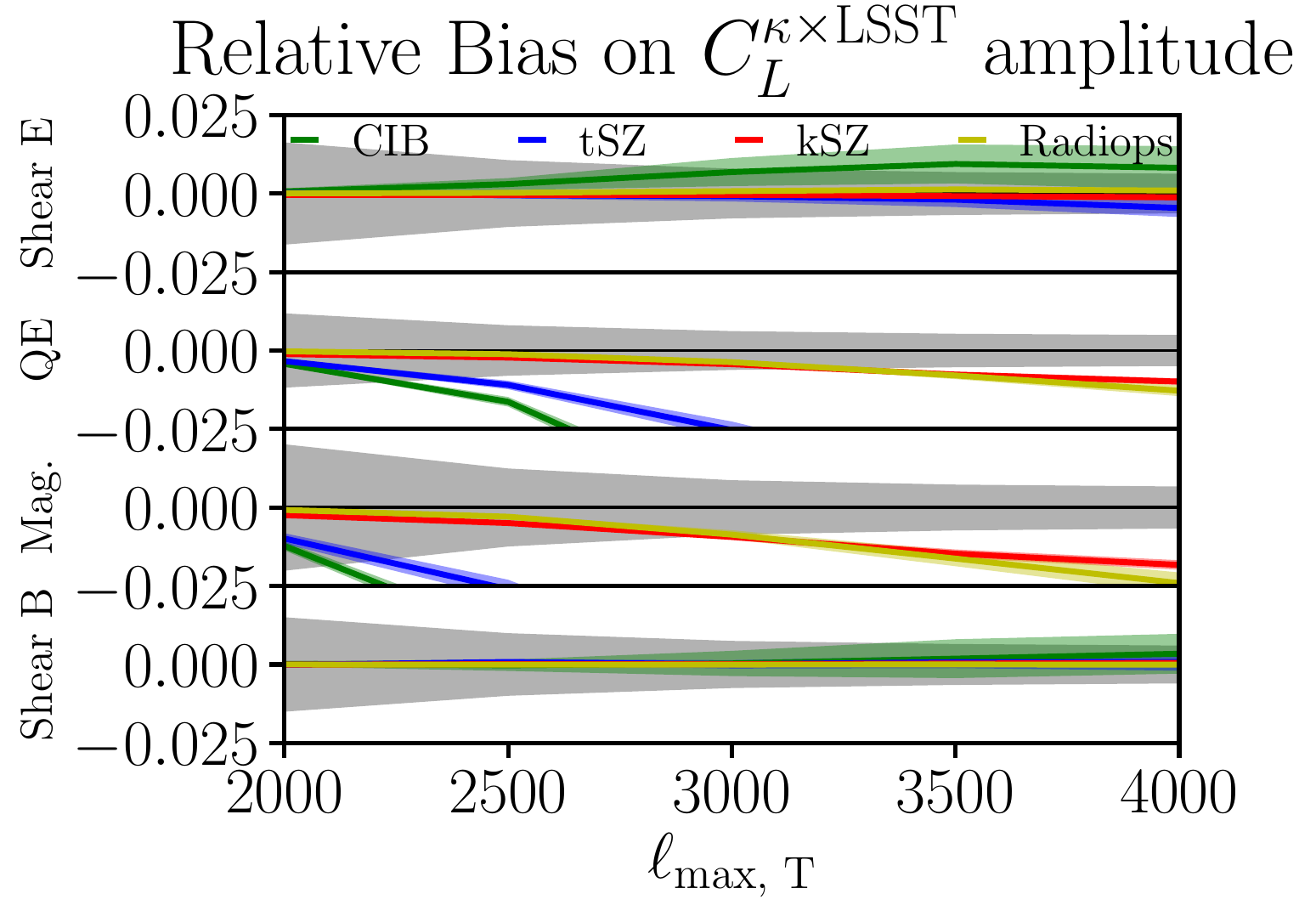}
\caption{
Relative bias on the amplitude of the CMB lensing power spectrum (left) and the amplitude of the cross-power spectrum of CMB lensing and LSST gold galaxies (right) due to the various foregrounds.
The grey band is the statistical error, including cosmic variance. 
This shows that the standard quadratic estimator is biased at the level of the statistical error bar for $\ell_\text{max, T} \simeq 2500$, compared to $3500$ for the shear E estimator.
Furthermore, the foreground biases for shear E and shear B have the same expectation value (therefore they are very similar on a realization by realization basis), so the combination shear E - shear B produces an unbiased estimator out to even higher $\ell_\text{max, T}$.
}
\label{fig:summary_bias}
\end{figure}

We show the bias on the amplitude of the lensing power spectrum and the amplitude of the cross-power spectrum of CMB lensing and LSST gold galaxies in Fig~\ref{fig:summary_bias}.
We have assumed that the Gaussian foreground contributions to the $N^0$ could be subtracted exactly, since this can be done from the measured power spectrum of the temperature map.
Considering the lensing auto-spectrum, the foreground biases equal the statistical uncertainty (including cosmic variance) for $\ell_{\text{max}, T}=2500$ for the QE and magnification,
compared to  $\ell_{\text{max}, T}=3500$ for the shear estimator.
This increase in $\ell_{\text{max}, T}$ has important implications in terms of lensing signal-to-noise, as described in the main text.

\end{document}